\documentclass[floatfix,superscriptaddress,twocolumn,amssymb,amsmath,prb,aps]{revtex4-2}

\pdfoutput=1

\usepackage{graphicx}
\usepackage{bm}
\usepackage{physics}
\usepackage[mathlines]{lineno}
\usepackage{setspace}
\usepackage[utf8]{inputenc}
\usepackage[english]{babel}
\usepackage[T1]{fontenc}

\usepackage{textalpha}

\begin{document}

\title{Investigating Mechanisms of State Localization in Highly-Ionized Dense Plasmas}

\author{Thomas Gawne}
\email{thomas.gawne@physics.ox.ac.uk}
\affiliation{Department of Physics, Clarendon Laboratory, University of Oxford, Parks Road, Oxford OX1 3PU, UK}

\author{Thomas Campbell}
\affiliation{Department of Physics, Clarendon Laboratory, University of Oxford, Parks Road, Oxford OX1 3PU, UK}

\author{Alessandro Forte}
\affiliation{Department of Physics, Clarendon Laboratory, University of Oxford, Parks Road, Oxford OX1 3PU, UK}

\author{Patrick Hollebon}
\affiliation{Department of Physics, Clarendon Laboratory, University of Oxford, Parks Road, Oxford OX1 3PU, UK}

\author{Gabriel Perez-Callejo}
\affiliation{Departamento de Física Teórica, Atómica y Óptica, Universidad de Valladolid, Valladolid, Spain}

\author{Oliver Humphries}
\affiliation{European XFEL, Holzkoppel 4, 22869 Schenefeld, Germany}

\author{Oliver Karnbach}
\affiliation{Department of Physics, Clarendon Laboratory, University of Oxford, Parks Road, Oxford OX1 3PU, UK}

\author{Muhammad F. Kasim}
\affiliation{Department of Physics, Clarendon Laboratory, University of Oxford, Parks Road, Oxford OX1 3PU, UK}

\author{Thomas R. Preston}
\affiliation{European XFEL, Holzkoppel 4, 22869 Schenefeld, Germany}

\author{Hae Ja Lee}
\affiliation{SLAC National Accelerator Laboratory, 2575 Sand Hill Rd, Menlo Park, CA 94025, USA}

\author{Alan Miscampbell}
\affiliation{Department of Physics, Clarendon Laboratory, University of Oxford, Parks Road, Oxford OX1 3PU, UK}

\author{Quincy Y. van den Berg}
\affiliation{Department of Physics, Clarendon Laboratory, University of Oxford, Parks Road, Oxford OX1 3PU, UK}

\author{Bob Nagler}
\affiliation{SLAC National Accelerator Laboratory, 2575 Sand Hill Rd, Menlo Park, CA 94025, USA}

\author{Shenyuan Ren}
\affiliation{Department of Physics, Clarendon Laboratory, University of Oxford, Parks Road, Oxford OX1 3PU, UK}

\author{Ryan B. Royle}
\affiliation{Department of Physics, Clarendon Laboratory, University of Oxford, Parks Road, Oxford OX1 3PU, UK}

\author{Justin S. Wark}
\affiliation{Department of Physics, Clarendon Laboratory, University of Oxford, Parks Road, Oxford OX1 3PU, UK}

\author{Sam M. Vinko}
\affiliation{Department of Physics, Clarendon Laboratory, University of Oxford, Parks Road, Oxford OX1 3PU, UK}
\affiliation{Central Laser Facility, STFC Rutherford Appleton Laboratory, Didcot OX11 0QX, UK}

\date{\today}

\begin{abstract}
We present the first experimental observation of K$_{\beta}$ emission from highly charged Mg ions at solid density, driven by intense x-rays from a free electron laser.
The presence of K$_{\beta}$ emission indicates the $n=3$ atomic shell is relocalized for high charge states, providing an upper constraint on the depression of the ionization potential. We explore the process of state relocalization in dense plasmas from first principles using finite-temperature density functional theory alongside a wavefunction localization metric, and find excellent agreement with experimental results.
\end{abstract}

\maketitle

\section{\label{sec:intro}Introduction}
Continuum lowering (CL) is a fundamental process in dense plasmas. When an atom is immersed in a plasma, electrostatic interactions between the atom and the plasma particles cause the continuum level of the atom to lower~\cite{ZIMMERMAN1980517,griem_1997,stewart1966lowering,ecker1963lowering}. This reduces all atomic binding energies, and can result in ionization when the amount of CL exceeds the binding energy of an electron. Continuum lowering has a direct impact on many important plasma processes, including the ionization and ion charge state distribution, the equation of state, the opacity, and transport properties. A good understanding of the physics behind the CL process is therefore vital to our ability to predictively model high energy density (HED) systems, including those relevant to inertial confinement fusion research~\cite{hu2011first} and to astrophysical plasmas~\cite{iglesias1991opacities}.

Collisional-radiative atomic kinetics models are widely used to understand the behaviour of HED plasmas, and constitute a key simulation framework needed to interpret the results of x-ray spectroscopy experiments. Because these models essentially deal with time-dependent ionization dynamics, they require clear definitions of ionization, and, in particular, of whether an electron is bound to some particular ion or not. Unbound electrons are considered free, and form a homogeneous (and often classical) free-electron gas.
In such simulations the process of CL is commonly accounted for via an ionization potential depression (IPD) model. The IPD is an energy that is subtracted from the electron binding energies in all ionic charge states. It depends on the electron density, temperature, and plasma ionization. As the name suggests, its effect is to decrease the amount of energy required to ionize an ion.
Ionization potential depression is a crude representation of an otherwise complex quantum manybody problem, but it is chosen because of its computational convenience and simplicity. For relatively hot systems, where IPD energies are small compared with other energies in the system, this approach can be adequate to represent the main effect of increasing electron density on the energetics of the plasma. In contrast, IPD models tend to falter for dense systems at lower temperatures, in particular for conditions where the mean particle separation distance starts to become comparable to the plasma Debye length, such as in warm dense matter.

There are many simple IPD models in use today. In the low-temperature and high-density limit, the ion-sphere (IS) IPD model is often used, the energy of which is evaluated from the spatially uniform free-electron density inside a Wigner-Seitz sphere~\cite{stewart1966lowering,ZIMMERMAN1980517,griem_1997}. At the opposite end, at high-temperatures and low-densities, Debye-H\"uckel (DH) theory~\cite{griem_1997} is thought to be applicable and the IPD is evaluated in terms of the Debye screening length. At intermediate conditions the Stewart-Pyatt (SP) model~\cite{stewart1966lowering} is popular, while yielding the IS and DH expressions at their respective limits. A modified version of the Ecker-Kröll (EK) model~\cite{ecker1963lowering,preston2013effects} has enjoyed some renewed popularity recently due to experimental results from the Linear Coherent Light Source (LCLS) \cite{SXR_LCLS} free-electron laser (FEL) at SLAC, but it has only been applied successfully to relatively low-Z materials.

\begin{figure}
    \centering
    \includegraphics[width=0.48\textwidth,keepaspectratio]{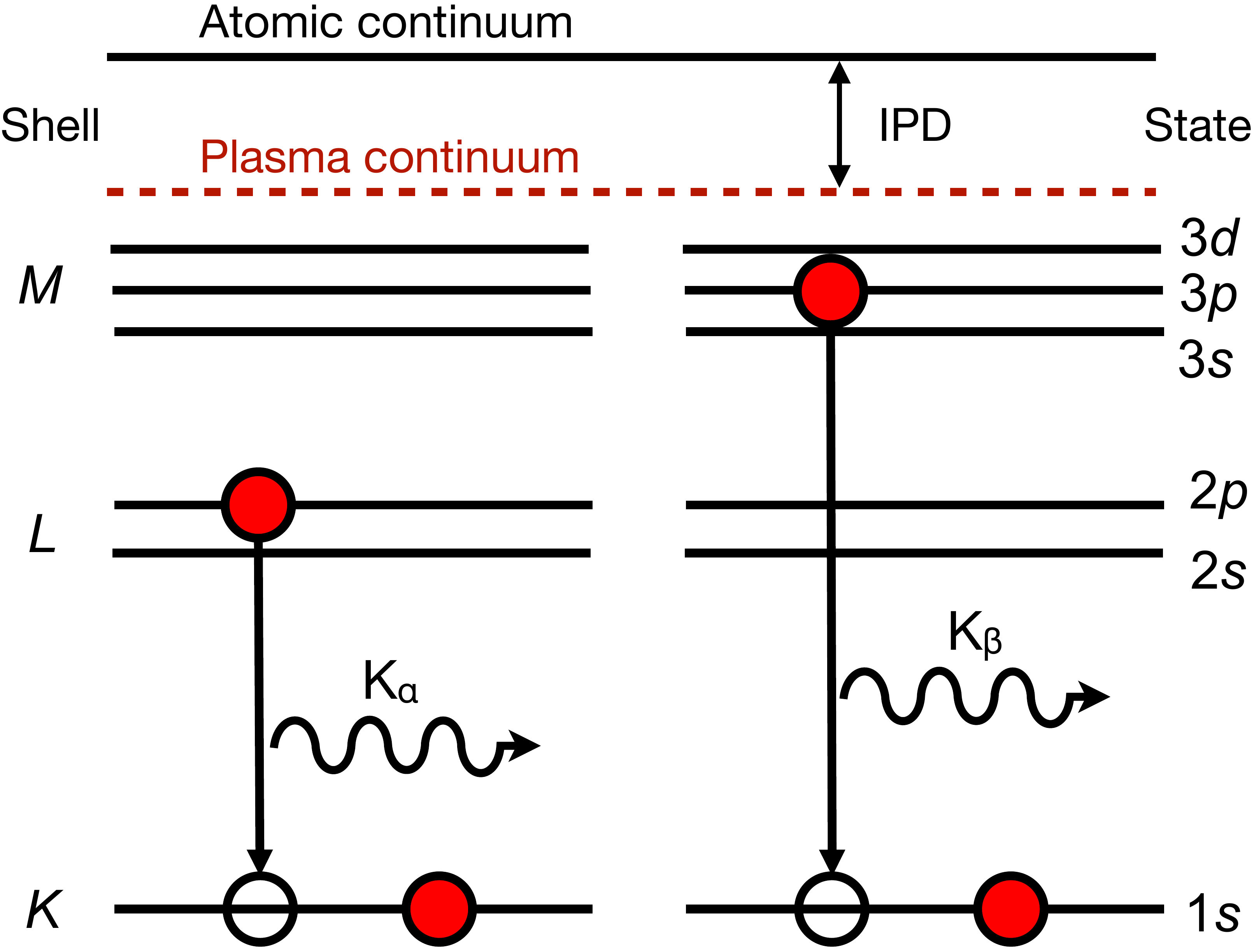}
    \caption{A schematic of K, L and M atomic shells, continuum level, IPD, and related K$_{\alpha}$ and K$_{\beta}$ transitions, discussed in the text. An additional hole in the K-shell would give rise to a `double-K hole' transition.}
    \label{fig:K Transition}
\end{figure}

In recent years there has been a series of novel experiments that attempted to measure the IPD in highly-ionized dense plasmas, and compare the experimental results with theoretical predictions. Much of this work relies on the physics of atomic $n=1,2,3$ orbitals (spectroscopically called the K-, L- and M-shells), and the related K$_{\alpha}$ and K$_{\beta}$ line emission. As we will use this nomenclature often, we illustrate the various terms using a simple atomic physics picture in Fig.~(\ref{fig:K Transition}).

An early experiment~\cite{vinko2012creation,Ciricosta2013PRL} at the LCLS measured the K-edge energy for the first few charge states in solid density Al plasmas containing cold ions and electrons heated up to 180 eV.
With a FEL pulse duration of 80 fs, the ions remain at low temperature throughout the duration of the pulse and resulting K$_{\alpha}$ emission due to the ion temperature evolution being dominated by electron-phonon interactions, which have timescales of order several picoseconds~\cite{Ng1995,Matthieu2011,White2014}. This allowed for CL to be examined at precisely known densities.
The experimentally inferred IPD showed good agreement with the EK model~\cite{preston2013effects}, and poor overall agreement with the SP model. Subsequent measurements by Ciricosta {\it et al.}~\cite{Ciricosta2016-nx} extended this technique to measure the onset of K$_{\alpha}$ emission in Mg, Al, Si, and various compounds of these materials. While the EK model showed better consistency with the data, the overall IPD of the full suite of materials could not be reproduced with any single IPD model. Importantly, the authors reported they were unable to detect any K$_{\beta}$ emission from higher charge states, concluding that the M-shell of the systems investigated was likely pressure ionized due to high IPD.

The IPD of high-lying charge states was, in parallel, explored in experiments~\cite{Hoarty2013,HOARTY2013661} conducted at the Orion laser facility~\cite{hopps2011overview}, where Al plasmas were driven to much higher temperatures of 500-700 eV, and shock-compressed up to around 3 times solid density. In contrast to the LCLS experiments K$_{\beta}$ emission was observed, and was used to diagnose the plasma conditions. In particular, the Ly$_{\beta}$ and He$_{\beta}$ lines were observed to vanish only at the highest densities achievable, a result attributed to the pressure ionization of the $n=3$ atomic shell. Compared with where the lines vanished in the experiment, the EK model substantially over-predicted the IPD for the reported experimental conditions, while the SP model under-predicted it. This indicates that the true value of CL should lie somewhere between the two, a finding consistent with more detailed theoretical predictions by Crowley~\cite{crowley2014continuum}.

Warm dense matter experiments conducted at higher densities and lower temperatures have produced complementary results to the experiments above.
Measurements~\cite{hansen2017changes} of shifts in the K$_{\alpha}$ and K$_{\beta}$ lines in Fe impurities in Be liners at Sandia's Z machine showed poor agreement with both the EK and SP models. The plasma temperature reached 10 eV, and the density reached 8 times solid density (electron density $n_{e} \simeq 2 \cross 10^{24}$ cm$^{-3}$).
In work at the OMEGA Laser Facility \cite{soures1996role}, Fletcher {\it et al.}~\cite{fletcher2014observations} compressed CH to around 7 times solid density and heated it to 10-20~eV temperatures. They find some agreement with the SP model, but in this regime the differences between EK and SP models are negligible.
However, similar experiments with CH capsules at the National Ignition Facility \cite{moses2009national} using spherically convergent shocks~\cite{kraus2016x} reaching higher temperatures of around 90~eV showed that the SP, IS and EK models all likely under-predict the experimental IPD.
More recently, measurements~\cite{kritcher2020measurement} of the equation of state along the principal shock Hugoniot of a hydrocarbon ($\rm{C_{9}H_{10}}$) showed an increase in the compressibility due to the partial ionization of the C core orbitals. Theoretical models that include electronic shell structure were better able to reproduce the measured EOS than those that lacked detailed structure, like the simple IPD models described above.

This rich collection of experimentally observed discrepancies has spurred a flurry of theoretical investigations into how best to model, predict, and understand, the effect of IPD.
Efforts have been made using ion-sphere models~\cite{iglesias2013fluctuations,rosmej2018ionization,zeng2022ionization}; Hartree-Fock-Slater calculations~\cite{son2014quantum}; models based on plasma theory~\cite{crowley2014continuum,rosmej2018ionization}; classical molecular dynamic simulations~\cite{calisti2015ionization}; quantum statistical models~\cite{lin2017ionization,lin2019quantum,ropke2019ionization}; Monte Carlo methods~\cite{stransky2016monte}; and density functional theory calculations with and without molecular dynamics~\cite{Vinko2014-fs,hu2017continuum,bethkenhagen2020carbon}.
Unfortunately, no consistent agreement has emerged between all these methods, but the majority of them do predict, as does the experimental data, that the IPD should lie energetically somewhere between the predictions of the SP and EK models. In this work we seek to resolve some of the existing discrepancies across the various theoretical modelling approaches, and present a first principles understanding of continuum lowering based on the localization and delocalization of states near the continuum of a dense plasma.

We start by presenting experimental results showing how M-shell states localize in highly ionized Mg, resulting in detectable K$_{\beta}$ line emission from high charge states including He$_{\beta}$. These emission lines are collected at exactly solid density, with electron temperatures of order 100~eV.
We then develop a theory that allows us to quantify whether a state is bound or free within the framework of finite-temperature density functional theory (DFT), and introduce the inverse participation ratio and a dimensionality parameter to measure its {\it boundness}. We provide physical insight into these methods using a few illustrative systems. We then apply these techniques to DFT simulations of systems in our experimental conditions, and compare the predictions with experimental observations. Finally, we use our method to extract a first-principles IPD calculation for Mg and Al, and compare it with simple analytical models. We also address some remaining discrepancies between our work and the earlier work on Mg by Ciricosta {\it et al.}~\cite{Ciricosta2016-nx}.

\section{Experimental measurements}

Isochoric heating experiments at x-ray FELs have shown that it is possible to heat solid-density systems to temperatures approaching 200~eV on femtosecond scales~\cite{Vinko2015,preston2017measurements,van2018clocking}, and probe their excitation thresholds using x-ray emission spectroscopy~\cite{vinko2012creation,Vinko:2015,Ciricosta:2016}. This method was successfully applied to the study of ionization thresholds in Al, Mg and other systems, from which the behaviour of the IPD could be studied as a function of ionic charge state~\cite{Ciricosta2013PRL,Ciricosta2016-nx}. Interestingly, while copious amounts of K$_{\alpha}$ emission is promptly observed from both single-core-hole (one vacancy in the $1s$ orbital) and double-core-hole states (two vacancies in the $1s$ orbital), these experiments found no radiative recombination from the M-shell in excited ions in the form of K$_{\beta}$ emission. Such emission is routinely observed in laser-plasma experiments, and is expected if the M-shell is bound and populated in low-Z simple metals like Al and Mg.

\begin{figure}
    \centering
    \includegraphics[width=0.48\textwidth,keepaspectratio]{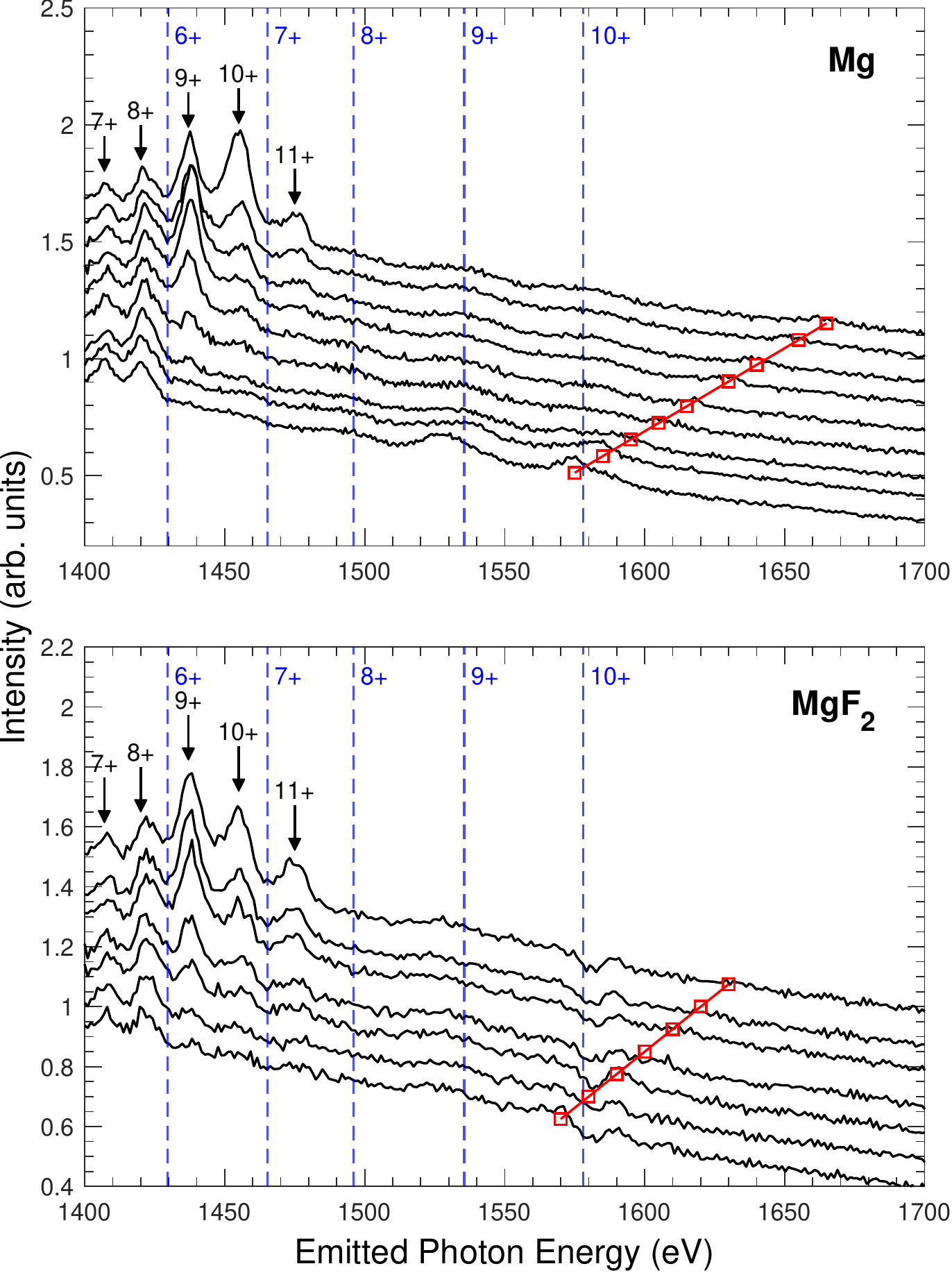}
    \caption{Spectra of Mg and MgF$_2$ for different incoming FEL photon energies. The intensity of the spectra has been normalized and displaced vertically for clarity. The black arrows indicate the double-K hole K$_{\alpha}$ lines. The energy of the atomic K$_{\beta}$ transitions calculated using AAEXCITE~\cite{peek1977continuum,cowan1981theory,aaexcite} are indicated by the vertical blue dashed lines. The labels refer to the charge state of the transition line. For these spectra, there are broad peaks corresponding to the K$_{\beta}$ transition lines for 8+, 9+ and 10+ Mg ions. The red open squares indicate the incoming photon energy for each spectrum; in ascending order for each material, these are: \\
    - Mg: 1575~eV, 1585~eV, 1595~eV, 1605~eV, 1615~eV, 1630~eV, 1640~eV, 1655~eV, 1665~eV.\\
    - MgF$_2$: 1570~eV, 1580~eV, 1590~eV, 1600~eV, 1610~eV, 1620~eV, 1630~eV. 
    }
    \label{fig:Mg MgF2 Spec}
\end{figure}

To investigate the presence of a bound M-shell we performed an experiment at the SXR endstation of the Linac Coherent Light Source (LCLS) x-ray FEL, operating in SASE mode. Thin foil samples of Mg and MgF$_2$ were irradiated with intense x-rays at a range of photon energies between 1570-1670~eV.
The samples used where 1~$\mu$m thick Mg foils deposited on a 10~$\mu$m Parylene N substrate, and coated with 30~nm Si$_3$N$_4$ to prevent sample deterioration and oxidation. The MgF$_2$ samples were similar, but without the Si$_3$N$_4$ coating. The emission spectrum was collected using a flat Beryl (10$\bar 1$0) natural crystal coupled to a charge-coupled device detector, and integrated over 50 shots, providing a resolution better than 0.4~eV within a spectral window between 1400--1700~eV.
Pulses of x-rays operating with a nominal duration of 60~fs, a spectral bandwidth of 0.4~\% FWHM, and an energy of 1~mJ were focused on target to spot sizes on the order of 10~$\mu$m$^2$ by means of a Kirkpatrick-Baez mirror~\cite{kirkpatrick1948formation}, yielding peak intensities reaching $10^{17}$~Wcm$^{-2}$. The focus was optimized {\it in situ} so as to maximize the double-core-hole emission from the Mg sample.
These parameters, and the experimental setup, are broadly similar to previous isochoric heating experiments at the LCLS, and further details can be found in ref.~\cite{Ciricosta2016-nx}. As was observed in previous work, such plasmas created on femtosecond timescales rapidly reach partial local thermodynamic equilibrium, and the observed emission spectrum is a good indicator of the plasma environment, containing hot thermal electrons and cold ions still in their lattice structure~\cite{Ciricosta:2016}. It was recently reported that for mid-Z metals such as Mg, non-thermal electrons remain unimportant in determining the short-time plasma kinetics and electron dynamics unless intensities exceed $10^{17}$~Wcm$^{-2}$, or for x-ray pulse durations below 10 fs~\cite{Ren:2023}.

The x-ray photon energies were chosen to allow us to resonantly pump the Mg$^{10+}$ K$_{\beta}$ (He$_{\beta}$) transition at the lowest energies, and to ensure we were able to drive the system above all single-core-hole K-edges at the highest energy. Copious amounts of Mg$^{10+}$ K$_{\alpha}$ (He$_{\alpha}$) emission was observed, indicating that high intensities were reached in the experimental setup, and demonstrating the successful creation of a substantial number of highly charged ions. This is consistent with previous measurements on Mg from ref.~\cite{Ciricosta2016-nx}. However, unlike in previous work, we focused here on the spectral window beyond the K$_{\alpha}$ emission from double core holes that terminate with the Mg$^{11+}$ K$_{\alpha}$ (Ly$_{\alpha}$) peak at around 1475~eV. We plot the measured emission from Mg and from MgF$_2$ in Fig.~(\ref{fig:Mg MgF2 Spec}). The red open squares in the figures indicates the resonance condition, where the photon energy of incident x-rays equals the emitted photon energy. On resonance, a small but noticeable quasi-elastic scattering peak is seen for the Mg sample, but not for MgF$_2$. The spectra are normalized and offset vertically for clarity.

At the lower emitted photon energy range, the spectra show a rich forest of double-core-hole K$_{\alpha}$ emission peaks indicated with black arrows. These correspond to spectral emission from bound-bound transitions of the group K$^0$ L$^x$ $\rightarrow$~K$^1$ L$^{x-1}$. The charge state is indicated using the number of holes in the K- and L-shells, but ignoring higher lying shells to simplify the notation. This means that 2+ is the lowest meaningful charge state of Mg that can be produced as the $3s$ electrons delocalize into the continuum. The 10+ transition is thus a combination of the $1s^0 2p^{2} \rightarrow 1s^1 2p^{1}$ and $1s^0 2s^1 2p^1 \rightarrow 1s^1 2s^1 2p^0$ transitions in a Mg ion with a He-like core, together with any possible satellite lines containing spectator electrons in the M-shell. At energies above the Ly$_{\alpha}$ line we observe two broad peaks in all the Mg spectra, associated with single-core-hole K$_{\beta}$ emission (K$^1$ L$^x$ M$^y$ $\rightarrow$ K$^2$~L$^x$~M$^{y-1}$, $y=1$ for no spectators) for the Mg$^{9+}$ and Mg$^{10+}$ charge states. In the Mg spectra for incoming photon energies of 1575-1605~eV, there is another broad peak corresponding to the Mg$^{8+}$ K$_{\beta}$ emission, though it appears to vanish at higher incoming photon energies.
The energies of these peaks are similar to the equivalent atomic K$_{\beta}$ transitions, calculated using the average approximation code AAEXCITE~\cite{peek1977continuum,cowan1981theory,aaexcite}, which are indicated by the vertical blue-dashed lines.
As with the double-core-hole K$_{\alpha}$ transitions, the charge state is indicated using the number of holes in the K- and L-shells in the lower state. The K$_{\beta}$ transition lines will therefore consist of different configurations L-shell configurations -- such as $1s^1 2p^1 3p^1 \rightarrow 1s^2 2p^1$ and $1s^1 2s^1 3p^1 \rightarrow 1s^2 2s^1$ for the 9+ transition -- together with satellite lines containing M-shell spectator electrons. 
For clarity, we indicate only the lowest energy atomic K$_{\beta}$ transitions in Fig.~(\ref{fig:Mg MgF2 Spec}) with no spectator electrons.

We note the presence of a dip in the spectrum for MgF$_2$ just above the He$_{\beta}$ energy –- this is an artefact due to the x-ray crystal used in the spectrometer. A similar feature is seen for the Mg data (not shown here) with the same crystal, but was no longer present in the data once the crystal was changed. While we were able to collect new spectroscopy measurements for Mg, the same was not possible for MgF$_2$ due to experimental time constraints. This defect in the spectra makes establishing the presence of He$_{\beta}$ emission challenging in MgF$_2$, and in general the presence of K$_{\beta}$ emission is difficult to determine unambiguously for this material. However, at least three K$_{\beta}$ emission lines can clearly be observed in Mg.

The K$_{\beta}$ emission is seen to be extremely weak, even when compared with the intensity of the Ly$_{\alpha}$ line, and is far weaker than the K$_{\alpha}$ emission from the same sample.
This may explain why previous experiments such as ref.~\cite{Ciricosta2016-nx} failed to observe any emission in this region, even when attempting to resonantly pump the lines with high intensity x-rays. The presence of K$_{\beta}$ emission is important in the context of IPD because it indicates a relocalization of the M-shell has taken place, and thus allows theoretical predictions to be compared directly with experiment. 
For example, in the context of simple models used in time-dependent atomic kinetics modelling, the SP IPD model predicts the M-shell to be localized for most charge states, while according to the EK model the M-shell is always pressure ionized. While it is well-documented via the shape of the K$_{\alpha}$ emission spectrum that the SP model underestimates the IPD in x-ray driven solid-density systems~\cite{vinko2012creation,Ciricosta2013PRL,Ciricosta:2016}, our observation of K$_{\beta}$ emission in Mg indicates that the EK model is no longer accurate for high charge states. This suggests a need to review the interpretation of experimental results from ref.~\cite{Ciricosta2016-nx}, but also motivates a more detailed examination of the process of M-shell relocalization in solid-density plasmas from a theoretical standpoint.

\section{\label{sec:Localization}Theory of state localization}

\subsection{\label{sec:SolidLocalization}Boundness and Localization}

Atomic kinetics simulations and IPD models require a binary picture of boundness. Electronic states are either purely bound or purely free, but this is not a complete picture of reality, especially in the hot dense plasma regime. However, the binary picture is convenient, in that it is conceptually simple and it allows for calculations of plasma conditions, opacity, and spectral emission to be performed quickly and efficiently. Recently, it has been shown that the ionization of a system -- which is a measure of how many electrons are bound to an ion -- can be related to the system's total pressure and the ion-ion pair distribution function~\cite{clerouin2022equivalence}, which provides a tangible connection between first-principles calculations and the binary picture. However, as this method deals with bulk properties, it does not concern itself with which states are bound or free, which is central to predicting spectra and opacity.

Defining rigorously the boundness of a state is a non-trivial problem. However, the following is expected: a purely free state should be spatially extended throughout a plasma, whereas one bound to an ion should be localized to some specific ion. The boundness of a state may therefore be associated to its spatial localization. Within this work, spatial localization is used to quantify the boundness of valence electronic states. By doing so, localization mechanisms can be investigated in a variety of scenarios, such as heating and compression. It also will allow us to scrutinize the basis of IPD models, as well as better justify which states are considered bound or free in these models. The problem is then how to define the spatial localization of a state. Typically, spatial localization is determined directly from the spatial extent of a state.  However, as we will discuss now, this is not a sufficient description for boundness. Instead, boundness should be associated with the response of the spatial extent of a state to changes in the unit cell. For simplicity, the arguments that are presented here be will be applied to simple unit cells. However, they are trivially applicable to any repeating cell, including those with less order requiring a larger collection of atoms to model, eg., a large supercell for a disordered warm dense matter plasma.


In order to quantify the localization of a state, the state itself must first be calculated at some level of theory. Here we choose to use finite-temperature DFT~\cite{HohenbergKohn,MerminGrandPot} to calculate the wavefunctions of the valence states, $\psi_{i}(\bm{r})$, through the Kohn-Sham (KS) equations \cite{KohnSham}:
\begin{equation}
    \label{eq:KohnSham}
    \left(-\frac{\hbar^{2}}{2m_{e}} \nabla^2 + \nu_{\rm{eff}}(\bm{r})\right)\psi_{i}(\bm{r}) = \epsilon_{i} \psi_{i}(\bm{r}),
\end{equation}
where $\nu_{\rm{eff}}$ is an effective potential containing the external potential, the Hartree interaction, and the exchange-correlation interaction terms, and $\epsilon_{i}$ is the KS energy of the $i$-th state. The electron density of the finite-temperature $N$-orbital system is:
\begin{equation}
    \label{eq:KS density}
    n(\bm{r}) = \sum_{i=1}^{N} f(\epsilon_{i}; \mu, T) \left| \psi_{i}(\bm{r}) \right|^2,
\end{equation}
where $f(\epsilon_{i}; \mu, T)$ is the occupation number of the state, given by the Fermi-Dirac equation, for a system at electron temperature $T$ with chemical potential $\mu$.
This approach is convenient for two reasons. Firstly, DFT makes no prior assumption on the boundness of individual Kohn-Sham states, and both valence and core states are treated on equal footing. Secondly, because this method allows us to calculate the electronic structure directly, any effect that would be considered to contribute to continuum lowering in the plasma system is already included self-consistently in the calculation of the ground state wavefunction. In taking this approach we make two assumptions. The first is that the Kohn-Sham wavefunctions are a good proxy for the electrons we wish to study, and the second is that our choice for exchange correlation functional in $\nu_{\rm eff}$ is sufficiently accurate to account for the many-body physics of our dense plasma.

\subsection{\label{sec:BlochLocalization}Localization of Bloch States}

The DFT calculations in this work were performed in both single primitive unit cells of crystals of the chosen materials and in larger disordered supercells.
In both cases, the electrons are treated with finite-temperature, and the ions are fixed in the crystal lattice. The difference is in the latter case, the ion positions are shifted from their perfect crystal lattice positions. For clarity, unit cells are the smallest repeating cell that can be infinitely repeated to describe a system -- even though supercells are used to describe a disordered system, in practice they are still infinitely repeating units of the system with a crystal symmetry that is pushed out to larger distances. Therefore, for both primitive cells and supercells, our KS states can be described by Bloch wavefunctions, $\psi_{b, \bm{k}}(\bm{r})$:
\begin{equation}
    \label{eq:BlochState}
    \psi_{b,\bm{k}}(\bm{r}) = u_{b,\bm{k}}(\bm{r}) e^{i \bm{k}\cdot\bm{r}},
\end{equation}
where $\bm{r}$ is the position vector, $b$ is the band index, $\bm{k}$ is the crystal momentum, and $u_{b,\bm{k}}(\bm{r})$ is a periodic function that has periodicity of the crystal lattice.
Bloch states are spatially extended: in a unit cell an arbitrary distance from the home unit cell, they have non-zero amplitude. Instead, localization is considered for the periodic function $u_{b,\bm{k}}(\bm{r})$ in a single unit cell. Spatially localized states -- such as an atomic orbital or bond -- are described by a $u_{b,\bm{k}}(\bm{r})$ that is localized within a region of the unit cell. Extended states have significant overlap with functions in other unit cells. Conceptually, this is similar to Kohn's theory of localization for insulating states \cite{kohn1964theory}, except here it is considered on a state-by-state basis rather than for the all-electron wavefunction. A Kohn-localized wavefunction $\ket{\psi}$ that is periodic in a lattice is one that can be decomposed into a sum of wavefunctions $\ket{\psi_{\bm{M}}}$ that are exponentially localized within disconnected regions $\bm{M}$, such as unit cells:
\begin{equation}
    \label{eq:KohnLocalization}
    \ket{\psi} = \sum_{\bm{M}} \ket{\psi_{\bm{M}}}.
\end{equation}
For localization, the overlap of $\ket{\psi_{\bm{M}}}$ and $\ket{\psi_{\bm{M'}}}$ is exponentially vanishing for different regions $\bm{M'} \neq \bm{M}$. Localization can therefore be considered within each unit cell rather than across the whole lattice.

Multiple schemes exist to describe the localization of Bloch states within the home unit cell. Typically, they involve measuring the size of a state either by quantifying its spread or by using a physical process, the strength of which depends on the degree of spatial localization. Two common schemes are the electron localization function (ELF) and Wannier functions.

The ELF~\cite{becke1990simple,savin1992electron} is a function in real space that quantifies regions in which electrons are localized by measuring Pauli repulsion in the space of the unit cell. The function then takes values $0 \le {\rm{ELF}}(\bm{r}) \le 1$, with the limits representing strongly delocalized to strongly localized, respectively. The ELF is particularly useful for identifying bonding features. However, the localization is only considered from the total density of the system, so it is not practical to calculate the localization of individual states.

The approach of Wannier functions~\cite{wannier1937structure} is to transform the Bloch states from extended states to functions that are localized.
Localization can then be determined by the changes in the spatial extent in the Wannier functions during a process~\cite{marzari2012maximally}. However, the transformation is gauge-invariant in the mixing of groups of connected bands. While one is free to choose a particular transformation, the interpretations drawn from a Wannier function, including its localization, can be very different depending on the choice of transformation, and no choice can be said to be better than another~\cite{marzari2012maximally}. 
While the freedom to choose the transformation scheme may be desirable for some applications, here it is not. Instead, a localization parameter that is not dependent on a chosen gauge is preferable.

Fortunately, an approach that does not suffer from the limitations described above can be found by computing inverse participation ratios and the effective dimensionality of the Kohn-Sham states. We will prefer this approach to study the rebinding of states, and describe it in more detail in what follows.

\subsection{\label{sec:IPR Localization}Inverse participation ratio localization}

Kohn's interpretation of localization says that well-localized states do not significantly cross the walls of the unit cell. This is a compelling picture. We note that this does not preclude localized wavefunctions that are spatially large in the unit cell; the only condition is that the overlap of two functions in neighbouring unit cells is exponentially vanishing. An important implication is that the spatial spread of a wavefunction does not provide a complete picture of localization. It therefore seems prudent to compare the spatial extent of wavefunctions that are extended or localized to see what information can be extracted about their localization.

\begin{figure}
    \centering
    \includegraphics[width=0.44\textwidth,keepaspectratio]{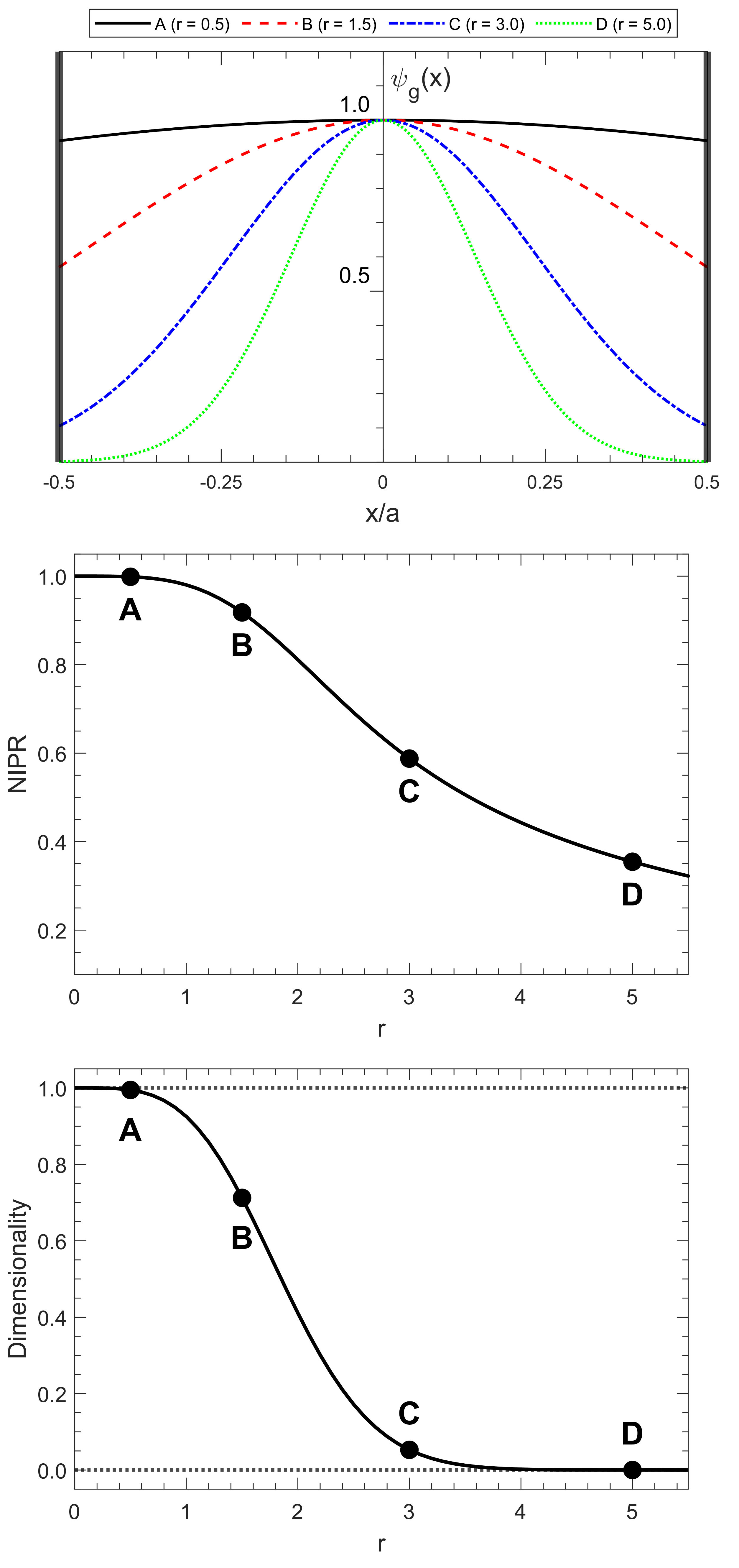}
    \caption{Plots of $\psi_{g}(x)$ (Eq.~(\ref{eq:Gaussian})), and its NIPR  (Eq.~(\ref{eq:IPR Delta Func})) and dimensionality (Eq.~(\ref{eq:DeltaFuncDim})) for different ratios of the lattice parameter to function size, $r = a/\sigma$. At $r=0.5$, the wavefunction fills much of the box, significantly crossing the walls of the unit cell. The dimensionality therefore has a value equal to the number of dimensions of the system ($D=1$). As $r$ increases, the function is increasingly localized within the cell, corresponding to a decrease in dimensionality. At $r>4.0$, $D=0$ and the functions are well-localized.}
    \label{fig:DeltaFuncDIM}
\end{figure}

Firstly, a measure of the spatial extent of the wavefunction is required. A simple way to do this is with the inverse participation ratio (IPR) \cite{bell1970atomic,wegner1980inverse}. For non-interacting states such as KS states the IPR is a measurement of extent by quantifying the portion of a space (real space, momentum space, number of atoms in a crystal, etc.) for which a wavefunction is significantly non-zero. For a wavefunction $\psi_{\alpha}(\bm{r})$ in a unit cell of volume $V$, the IPR is given by:
\begin{equation}
    \label{eq:IPR Def}
    \mathcal{I}[\psi_{\alpha}(\bm{r})]  = \\ 
    \frac{\left( \int_{V} \left| \psi_{\alpha}( \bm{r} ) \right|^2 \, d \mathbf{r} \right)^{2} }{\left(\int_{V} \left| \psi_{\alpha}( \bm{r} ) \right|^4 \, d \mathbf{r} \right)}.
\end{equation}
The IPR is a bounded quantity, taking values between $0 \le \mathcal{I} \le V$. In usual applications of the IPR, it describes the localization of a state, with the limits having simple interpretations: delocalized states, such as single plane waves ($\psi = V^{-1/2} e^{i\bm{G}\cdot\bm{r}}$), have $\mathcal{I} = V$. Well-localized states, such as a delta-function, have $\mathcal{I} \rightarrow 0$. In other words, localized states occupy a small volume of the cell while delocalized states fill most of it. This follows the same spatial extent interpretations of localization from the previous section. However, the size of the wavefunction does not give a complete picture of its localization. To illustrate this, consider two different model functions in an isolated one-dimensional cell with lattice parameter $a$.

First, consider a box function:
\begin{equation}
    \label{eq:Box Func}
    \psi_{\rm{box}}(x) = 
    \begin{cases}
         0 & \text{if } |x| > \gamma/2,\\
         A/2 & \text{if } |x| = \gamma/2,\\
         A & \text{if } |x|< \gamma/2,
     \end{cases}
\end{equation}
where the normalization constant $A = \gamma^{-1/2}$, and $\gamma$ is the width of the box ($\gamma>0$). For $\gamma < a$, $\psi_{\rm{box}}(x)$ is contained entirely within the cell; it does not cross the walls of the box. By Kohn's definition of localization, a repeating set of these functions in different unit cells would form a localized Bloch function. The IPR of this function is:
\begin{equation}
    \label{eq:IPR Box Func}
    \mathcal{I}[\psi_{\rm{box}}(x)] = \gamma,
\end{equation}
which is just the width of the box function. The function is localized and has a finite size that could occupy a large portion of the unit cell. Importantly, its IPR is independent of the lattice parameter. 

Next, consider the Gaussian function:
\begin{equation}
    \label{eq:Gaussian}
    \psi_{g}(x) = A e^{-x^2/\sigma^2},
\end{equation}
where $\sigma$ is the width of the function ($\sigma>0$) and $A$ is the normalization constant
\begin{equation}
    \label{eq:Finite Delta Norm}
    A^{-2} = \sigma \sqrt{\frac{\pi}{2}} \erf \left( \frac{a}{\sqrt{2}\sigma} \right),
\end{equation}
with $\erf(x)$ the error function~\cite{AndrewsLarryC1998Sfom}. In the limit that $\sigma \rightarrow 0$, $\psi_{g}(x)$ becomes a delta function, $\delta(x)$. This function is plotted in Fig.~(\ref{fig:DeltaFuncDIM}) for various ratios of $r = a/\sigma$.
For this function, the IPR is:
\begin{equation}
    \label{eq:IPR Delta Func}
     \mathcal{I}[\psi_{g}(\bm{x})] = \sigma \sqrt{\pi} \\
     \frac{ \erf(r/\sqrt{2})^2 }{ \erf(r) }.
\end{equation}
The IPR is dependent on the volume of the cell, but the dependence is non-trivial, and is not a binary picture of bound or free. The normalized IPR,  ${\rm{NIPR}}=\mathcal{I}/a$, is plotted in Fig.~(\ref{fig:DeltaFuncDIM}).
There are two extremes for the IPR of this function. The first is where $r \gg 1$  (i.e. $\sigma \ll a$, a well-localized function). In this case, $\erf(r) \rightarrow 1$, and $\mathcal{I} = \sigma \sqrt{\pi}$. Again the IPR can take a finite value, but is independent of the size of the cell. In the case of a delta-function ($\sigma \rightarrow 0$), the IPR is zero, as expected. The other case is when $r \ll 1$. This function is completely delocalized as it significantly crosses the walls of the unit cell. A Taylor expansion of the error function gives $\erf(r) = \frac{2}{\sqrt{\pi}}\left( r - \mathcal{O}(r^3) \right)$, which leads to an $\mathcal{I} = a$. The IPR of the completely delocalized state is directly proportional to the volume.

In general, well-localized states, which may be considered to be bound, have IPR values which are independent of the cell volume. Completely delocalized states have $\mathcal{I} \propto V$. The localization of a state can therefore be measured from its volume dependence.

\subsection{\label{sec:Dim Localization}Localization from Dimensionality}

To overcome the general complexity of the IPR's volume-dependence, consider instead the response of the IPR when the unit cell is expanded by a small amount. In real space, a $D$-dimensional unit cell is drawn by its set of $D$ lattice vectors, which are given by primitive lattice vectors, ${\bm{p}}_{i} = \sum_{j=1}^{D} p_{i,j} \hat{\bm{x}}_j$, and the set of lattice parameters, $a_{i}$. The set of lattice vectors are ${\bm{R}}_i = a_i \bm{p}_i$. The lattice vectors can be expressed in a matrix $({\bm{R}}_{i})_{j}$ which can be used to determine the volume of the unit cell:
\begin{equation}
    \label{eq:Volume Def}
    V = \det [ ({\bm{R}}_{i})_{j}] = \det[ a_i p_{i,j} ].
\end{equation}
Expanding the lattice parameters of the unit cell by a small amount $a_{i} \rightarrow a_{i} + \delta a_{i}$, the volume increases to $V' = \det[a_i(1 + {\delta a_{i}}/a_i) p_{i,j}]$. If all the lattice parameters change by the same proportion ${\delta a_{i}}/a_i = {\delta a}/a = \epsilon$ then the expanded volume can be expressed as:
\begin{equation}
    \label{eq:Volume Expansion}
    \begin{split}
        V' & = \det[ a_i (1+\epsilon) p_{i,j} ] \\
           & = (1+\epsilon)^D \det[ a_i p_{i,j} ] \\
           & = (1+\epsilon)^D V.
    \end{split}
\end{equation}
Now consider a fully delocalized state, which has $\mathcal{I} \propto V$. The IPR of the delocalized state in the expanded cell is $\mathcal{I}' = (1+\epsilon)^D \mathcal{I}$. If $\epsilon \ll 1$, then $\mathcal{I}' \approx (1+D\epsilon) \mathcal{I}$. The number of dimensions of the system, $D$, of the fully delocalized state is then given by:
\begin{equation}
    \label{eq:Dim Def}
    D = \frac{a}{{\delta a}} \frac{\mathcal{I}' - \mathcal{I}}{\mathcal{I}} = \frac{a}{\mathcal{I}} \frac{d \mathcal{I}}{da},
\end{equation}
As will be shown below, Eq. (\ref{eq:Dim Def}) can be used to measure the localization of a state. It is termed `dimensionality' as it gives the effective number of dimensions in which a function is delocalized.
For the three-dimensional unit cells used here, a fully delocalized state, such as a plane wave, will recover $D = 3$, indicating it is delocalized in all directions. By comparison, a delta-function has $D = 0$, a reflection of the fact that a delta-function is a zero-dimensional object, and it does not respond to the increasing volume of the system.
An exact, logarithmic definition for $D$ is clearly also a possibility. However, we will prefer the definition given in Eq.~(\ref{eq:Dim Def}) as it also allows for a simpler interpretation: a measurement of the rate at which a state {\it changes size} as the unit cell expands, with the unit cell having $D$ equal to the number of dimensions in the system. This interpretation will be conceptually clearer when the dimensionality is applied to more complex systems where $D < 0$ and $D > 3$ may occur.

The dimensionality has simple interpretations: if $D$ is large, the states are delocalized, with $D=3$ indicating complete delocalization. If $D$ is small the states are localized, with $D = 0$ indicating perfect localization. For a three-dimensional case, $D<1$ is typically a good indicator of localization as the function is not delocalized in any single dimension.
Additionally, states with $D < 0$ may occur, indicating `shrinking' as the unit cell expands, as well as states with $D>3$ indicating expansions faster than the unit cell expands.

\subsection{\label{sec:Model System Dim}Dimensionality of Model Functions}

Before discussing its application to real systems, it will be informative to apply the dimensionality to the model wavefunctions of Eqs.~(\ref{eq:Box Func}) and~(\ref{eq:Gaussian}), in a one-dimensional unit cell of size $a$.
For the box function, $\psi_{{\rm{box}}}(x)$, $\mathcal{I} = \gamma$, which is the width of the box. Importantly, the IPR is independent of the volume of the cell. The dimensionality of this function is therefore $D = 0$. This reflects the fact that expanding the box does not change the IPR of the function, and so the state is localized.

The dimensionality of $\psi_{g}(x)$ is a smooth function that, like the IPR, depends on the relative sizes of the unit cell to the function, $r$:
\begin{equation}
    \label{eq:DeltaFuncDim}
    D = \frac{2r}{\sqrt{\pi}} e^{-r^2} \left[\sqrt{2}e^{-r^2 / 2} \erf \left(\frac{r}{\sqrt{2}}\right)^{-1} - \erf \left(r \right)^{-1} \right].
\end{equation}
Like the IPR, the dependence on $r$ is non-trivial, but for the purpose of determining localization all that is important is single values of the dimensionality for a given $a$ and $\sigma$. Eq.~(\ref{eq:DeltaFuncDim}) is plotted in Fig.~(\ref{fig:DeltaFuncDIM}), along with the functions for various $r$ for comparison.
In the case of a localized $\psi_{g}(x)$ with $r \gg 1$, $\mathcal{I} \propto \sigma$. Again, the volume-independence of the IPR for a localized function leads to $D = 0$. But, when $r \ll 1$ and the state is delocalized, $\mathcal{I} = a$. The dimensionality of this state is then $D = 1$, recovering the number of dimensions of the system for a fully delocalized state.

To show how the dimensionality reveals the number of delocalized dimensions, we calculate the dimensionality of the function $\Psi(x,y) = \psi_{{\rm{box}}}(x) \psi_{g}(y)$ in a two-dimensional unit cell of size $a \cross b$. The width of $\psi_{{\rm{box}}}(x)$ is $\gamma$, and for $\psi_{g}(y)$ it is $\sigma$. Trivially, the $x$-direction will always be localized, while the localization of the $y$-direction will depend on $\sigma$.
In this simple case where the directions are decoupled from each other, $\mathcal{I}[\Psi(x,y)] = \mathcal{I}[\psi_{{\rm{box}}}(x)] \cross \mathcal{I}[\psi_{g}(y)]$. Therefore the total IPR of the function is:
\begin{equation}
    \label{eq:IPR Mix Func}
     \mathcal{I}[\Psi(x,y)] = \sqrt{\pi} \gamma \sigma  \\
     \frac{ \erf(b/{\sqrt{2}\sigma})^2 }{ \erf(b/\sigma) }.
\end{equation}
The IPR only depends on the lattice parameters through $\psi_{g}(y)$. The dimensionality of this function is therefore the same as Eq. (\ref{eq:DeltaFuncDim}). In the localized limit of $\sigma \ll b$, the dimensionality is $D = 0$ as $\Psi(x,y)$ is localized in all directions. It also reflects that $\Psi(x,y)$ does not change size when the unit cell expands. In the delocalized limit of $\sigma > b$, the dimensionality is $D = 1$, indicating that the state is only delocalized in the $y$-direction. Equivalently, we can say that $\Psi(x,y)$ is expanding at half the rate of the unit cell. For completeness, if the $x$-direction is also delocalized, then $D \rightarrow 2$.

\subsection{\label{sec:Dim Function Overlap}Effects of Function Overlap}

So far, the model functions considered have been single functions in isolated unit cells, such that there is no overlap of functions between different unit cells or within unit cells. But, except for very well-localized states, there will often be overlapping wavefunctions. We now examine the effects of function overlap, and the additional dimensionalities it allows for.

\begin{figure}
    \centering
    \includegraphics[width=0.44\textwidth,keepaspectratio]{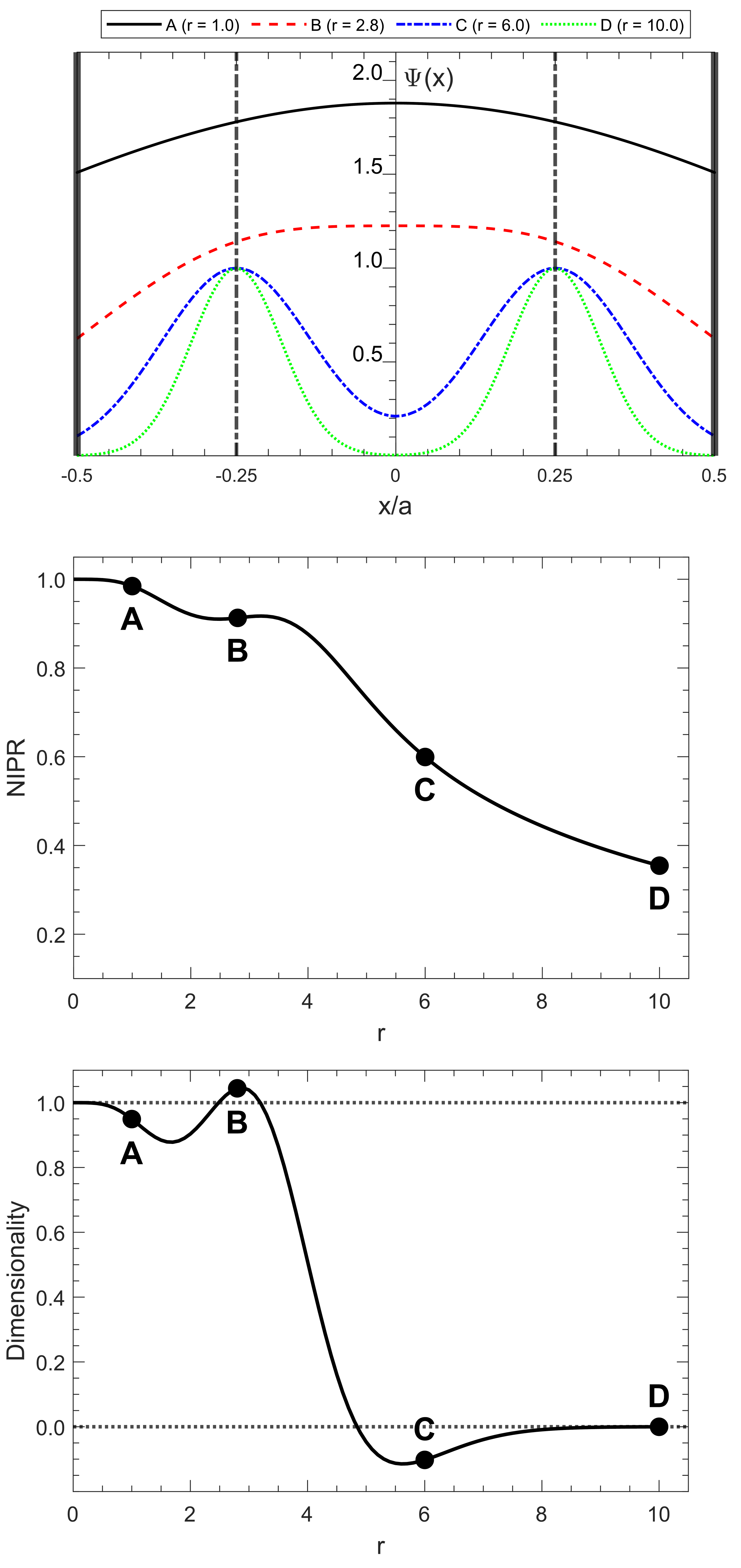}
    \caption{Plots of $\Psi(x)$ (Eq. (\ref{eq:Two Atom Psi})), and its NIPR and dimensionality for different ratios of the lattice parameter to function size, $r = a/\sigma$. $\sigma$ is the width of $\psi_{g}(x)$. The centers of the two orbitals are indicated by the vertical black dashed lines. Two interesting features stand out: at $r \ge 4.8$, $D<0$; and $D>1$ for $2.4 < r < 3.2$, during which the NIPR briefly increases.}
    \label{fig:2AtomFUNC}
\end{figure}

Consider the case of two functions in an isolated cell of size $a$, which has a total wavefunction:
\begin{equation}
    \label{eq:Two Atom Psi}
     \Psi(x) = \psi_{g}(x-a/4)+\psi_{g}(x+a/4),
\end{equation}
where $\psi_{g}(x)$ is the Gaussian from Eq.~(\ref{eq:Gaussian}). This situation is equivalent to two atoms in an isolated unit cell with a single orbital. Some examples are plotted in the top figure of Fig.~(\ref{fig:2AtomFUNC}) for different ratios $r = a/\sigma$, where $\sigma$ is the width of $\psi_{g}(x)$.
The IPR and dimensionality have an exact solution, but the functional forms are significantly more complex than those of Eqs.~(\ref{eq:IPR Delta Func}) and (\ref{eq:DeltaFuncDim}), so they are omitted here.
The NIPR and dimensionality of this function are also plotted in Fig.~(\ref{fig:2AtomFUNC}). Two interesting features appear in the dimensionality plot: first, the dimensionality is negative for $r \ge 4.8$, indicating a ``shrinking'' effect for the wavefunction. Second, the dimensionality goes above the number of dimensions in the system ($D>1$) for $2.4 < r < 3.2$, indicating the function is expanding faster than the unit cell.

The reason for this more complex localization behaviour can be understood in the overlap behaviour of the functions.
For $r=1.0$, when $D$ still indicates the function is delocalized, $\Psi(x)$ has a slight curvature that significantly crosses the walls of the cell, with a similar shape to that of Fig.~(\ref{fig:DeltaFuncDIM}) function A.
 At $r=10.0$, when $D=0$ and the function is well-localized, the two peaks are well-distinguished and have little overlap.
When $r=6.0$, the peaks are distinguished, but there is still a noticeable overlap between the functions. However, increasing $a$ reduces this overlap, which results in an effective shrinking of the IPR, hence the dimensionality is negative. Such dimensionalities represent a region of states relocalizing.
When $D>1$, such as at $r=2.8$, the function's behaviour is more complex. Initially, before $D>1$, the function is curved much like for $r=1.0$; as $r$ increases so does the curvature and the dimensionality decreases. However, at about $r=2.4$, the central part of the function flattens. This flat region grows until about $r=3.2$, at which point the two peaks begin to separate out. The flat region of the function is a delocalized region of the function. This results in the IPR growing with $a$ causing the dimensionality to increase after $r>2.0$. For a region $2.4 < r < 3.2$, the growth is faster than the rate of expansion of the unit cell, hence $D>1$.
Once the center of the peaks separates enough that the peaks begin to become distinguished, the IPR begins to decrease as $a$ increases. The functions therefore begin to relocalize again.

\subsection{\label{sec:Real System Dim}Dimensionality in Real Systems}

The behaviour of real systems is significantly more complex than the simple functions examined thus far. Not only do the wavefunctions take much more complex functional forms than the model cases, one must also contend with the fact that changing the lattice parameter also changes the solutions to the KS equations. Increasing the lattice parameter will change the potential the electrons interact with, as well as changing the density of the system. Therefore, changing the lattice parameter will affect the solutions to the KS equations and the minimization of the Mermin grand potential \cite{MerminGrandPot}. However, if the expansion is sufficiently small then differences will be negligible, and the average dimensionality should predominantly capture the response to expanding the unit cell. On the other hand, if the expansion is too small, then the computed wavefunctions will be too similar to determine the localization due to limits on computational accuracy. In this work, ${\delta a}/{a} = 0.1\%$ was found to be sufficient to ensure we capture the localization behaviour: the eigenvalues change by $< 0.5\%$ ($< 0.61$~eV absolute change) for all the materials, so differences in the systems are small; while states with known localizations, such as atomic L-shells and high energy plane waves, had the expected dimensionalities of $D < 1$ and $D = 3$ respectively, so changes are still meaningful.

It was found that $D>3$ tends to occur just above the continuum edge, which typically marks boundary between localized and delocalized states, or ``bound'' and ``free'' states in IPD models. The large dimensionality indicates these states are very sensitive to the change in conditions, which is not unexpected at the edge. From our model functions, this appears to be due to growing overlaps in functions that are delocalized. In testing with systems that had known localizations, $D<0$ was seen to occur with states that were expected to be localized.

When the dimensionality is calculated for the individual KS wavefunctions, there is large variation due to taking the numerical derivative of KS states, and because the derivative captures both the localization and changes in the KS system. However, the average of the dimensionality is found to be at the expected values, so it is the averaged dimensionality that we consider. The averaging is done using a moving mean with an energy window of 4 eV.

\section{\label{sec:methods}DFT Simulations}

We obtain the wavefunctions needed for our dimensionality study via finite-temperature Kohn-Sham DFT simulations using a locally-modified version of the plane wave code ABINIT v8.10.3~\cite{Gonze2016,Torrent2008,Bottin2008}. Modifications include the hybrid scheme detailed in Section~\ref{sec:KS-PWA}, a module to calculate the IPR of the real KS wavefunctions, and a finite-temperature local density approximation (LDA) exchange-correlation (xc) functional~\cite{FT-LDA} for testing if explicit finite-temperature effects on exchange-correlation are important.

To avoid spurious effects due to perfect symmetries in primitive cells we model our system with supercells: 32 atom ($2\times2\times2$) Al; 54 atom ($3\times3\times3$) Mg; and 48 atom ($2\times2\times2$) MgF$_2$. These sizes are kept relatively small to keep the calculations computationally manageable. The ions were evolved in a molecular dynamics simulation using ABINIT's isokinetic driver, with an ion temperature of 300 K in all cases, and an electron temperature $T_e = 1$~eV for Mg and Al, and $T_e = 300$ K for MgF$_2$; the specific electron temperatures are largely unimportant to the ion motion. The ions were allowed to evolve for a large number of time steps ($> 1000$) after reaching equilibrium to provide a selection of different ion positions over a few oscillations about the ions' origins. The ion positions therefore represent realistic deviations from the perfect crystal lattice that would be seen in experimental conditions. Static ground state calculations for a number of different ion positions with the desired electron temperature were then performed, and the dimensionality of the KS states averaged over the different ion positions.


For efficiency the ion cores are modelled in the PAW pseudopotential scheme~\cite{BlochlPAW}, with the $1s$ orbital frozen in the core. The potentials were all generated for a ground-state configuration at $T = 0$ K using the \textit{Atompaw} code~\cite{atompaw}. For the temperatures considered in this work the thermal depopulation of the $1s$ orbital is negligible, justifying this frozen core approximation. In general, PAW potentials generated at zero temperature can be applied in finite-temperature calculations provided the frozen populations are set appropriately~\cite{Hollebon_2021,PhysRevE.93.063207}.

The xc functional plays a minor role in this work, as we are mainly interested in studying the density of states (DOS) and dimensionality, which we found to have little dependence on the choice of xc model. For this, the PBE form of the generalized gradient approximation (GGA)~\cite{PBE-GGA} is known to perform well across multiple system, including those which we are studying here. Nevertheless, we performed a comparison of the DOS and the dimensionality for Mg and MgF$_2$ between the zero-temperature GGA-PBE and LDA-PW~\cite{LDA-PW} functionals, and the finite-temperature LDA functional GDSMFB~\cite{FT-LDA}, at a few high temperature conditions. The GDSMFB eigenvalues were slightly shifted compared with results from GGA-PBE and LDA-PW, however the chemical potential was also shifted by a similar amount. The ionizations, shape of the DOS, and the dimensionalities were the same in all cases.

\begin{figure}
    \centering
    \includegraphics[width=0.48\textwidth,keepaspectratio]{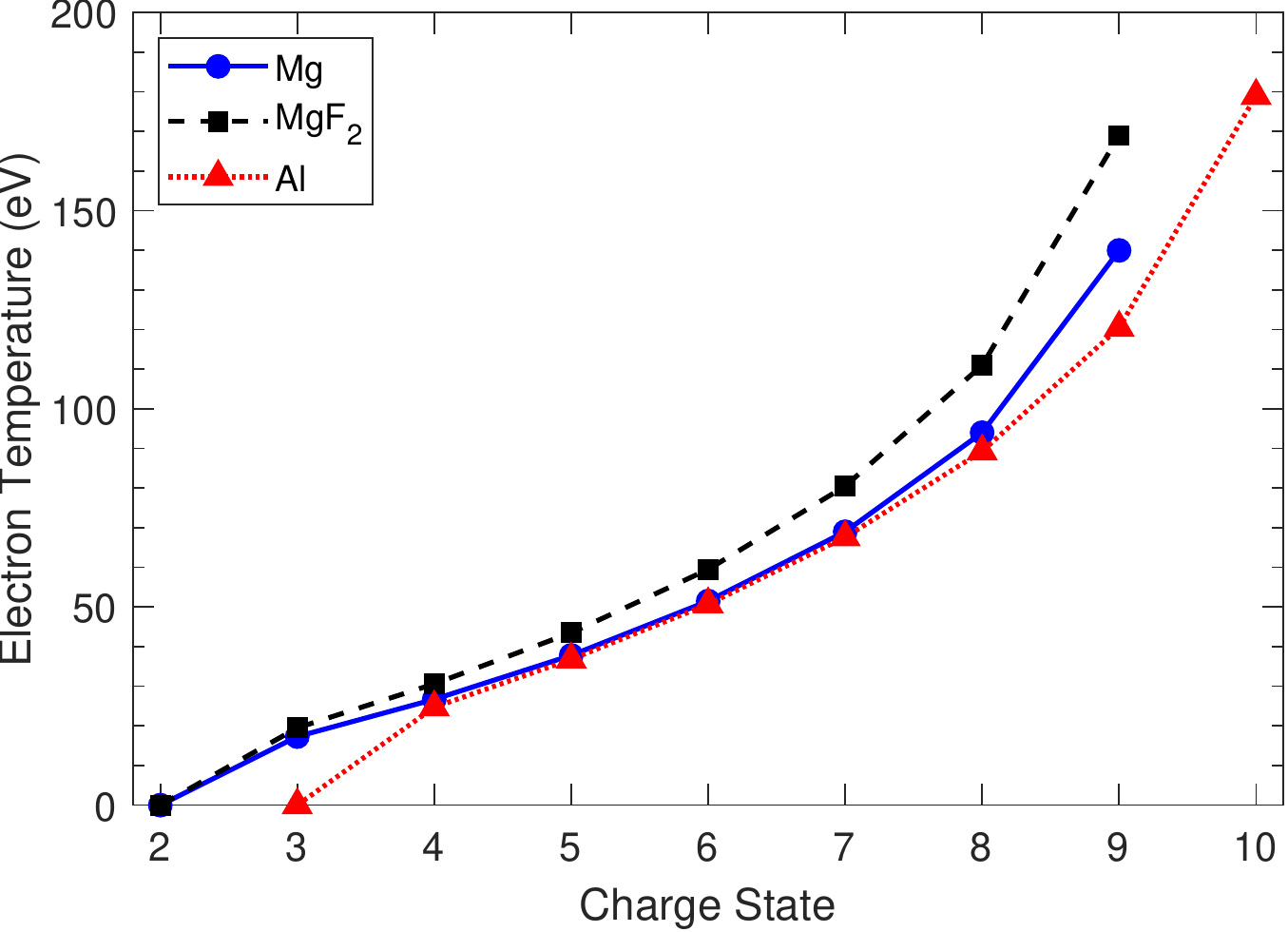}
    \caption{The temperatures required to produce each average charge state of the metal ion in Mg (blue circles), MgF$_2$ (black squares), and Al (red triangles). Note the ionization of 2 and 3 for Mg and Al in the $T\rightarrow0$ limit, consistent with a metallic ground state containing the full M-shell in the delocalized conduction band.}
    \label{fig:ChargeVsT}
\end{figure}

Our calculations are performed over a range of temperatures consistent with those reached in isochoric heating experiments, as determined by time-dependent atomic kinetics simulations. This enables a simple comparison with the experimental data, while at the same time allowing us to study how the dimensionality of the valence states changes as the temperature is increased. The temperatures for the three materials were chosen to thermally ionize integer numbers of electrons from the L-shells of the metal ions. We show the temperatures required to produce each average charge state of the metal ion in the DFT calculations in Fig.~(\ref{fig:ChargeVsT}).
The temperatures required to substantially thermally ionize a Mg ion are considerable, which poses fundamental challenges to KS DFT. To resolve this we have implemented the hybrid Kohn-Sham plane-wave-approximation scheme (KS-PWA) for high-temperature work by Zhang \textit{et al.}~\cite{Zhang2016}, which we outline briefly in the following section.

As we are primarily interested in the DOS and the dimensionality, simulation parameters were chosen to converge these properties. The dimensionality was more challenging to converge than the DOS, so once the dimensionality was converged, so was the DOS. Good convergence was considered to be reached once the dimensionality did not change substantially so that the localization of the states could be determined. In practice, this means that systems were converged until the maximum absolute differences in the dimensionality was $< 0.2$ and a mean absolute difference of $< 0.03$. The largest differences in the dimensionality were generally in small energy regions, and the localization of these regions could still be determined.
With the KS-PWA scheme, our static supercell calculations could be performed with just 2200 bands for Al and MgF$_2$, and 4400 bands for Mg, which gives good convergence of the calculations, and a DOS sampled up to sufficiently high energies for our study. The energy of the highest states is also well within the regime of the hybrid scheme~\cite{BLANCHET2022108215}, and the shape of the DOS and the states having $D=3$ confirmed we reached the plane wave limit.
As supercell calculations were performed, the DOS and dimensionality are averaged over a number of different ion positions until they are converged: 18 positions for Al and MgF$_2$, and 14 for Mg.
While the definitions of IPR and dimensionality do not explicitly consider the ion centers, the form of a wavefunction -- and so its eigenvalue for the DOS -- does depend on local density conditions, and experimentally it is an average over different local conditions that is measured.
Only the gamma point is sampled in the Brillouin zone (BZ) as the size of the BZ is greatly reduced by the disorder and large lattice parameters of the supercells. Furthermore, sampling just the gamma point excludes symmetry effects.
A plane wave cut-off energy of 25 Ha for sampling the PAW pseudo-wavefunctions, and 50 Ha for sampling the real wavefunctions, was found to be sufficient for converging for the dimensionality parameter and the eigenvalues.

\subsection{\label{sec:KS-PWA}Hybrid scheme for high temperatures}

Finite-temperature KS DFT simulations are typically limited to temperatures of a few 10s of eV. The reason for this lies in the Fermi-Dirac distribution, which gives the occupation of states calculated in DFT:
\begin{equation}
    \label{eq:FDocc}
    f(\epsilon; \mu, T) = \frac{1}{1 + {\rm exp}((\epsilon-\mu)/k_{B}T)},
\end{equation}
where $\epsilon$ is the energy of the state, $\mu$ is the chemical potential, and $T$ is the (electronic) temperature.
For temperatures up to around the Fermi temperature the number of states that need to be considered is relatively small as the occupation number quickly tends to zero at energies above the Fermi level.
However, as the temperature increases above the Fermi temperature the occupation number of states far above the Fermi level quickly becomes considerable. This causes two issues: firstly, a large number of states must be treated in the DFT simulation in order to perform an accurate calculation of the total energy; and secondly, the kinetic cut-off energy needs to be very high to access these states. Increasing the number of states and the kinetic cut-off energy quickly raises the computational cost of performing finite-temperature calculations.

Previous experimental work at x-ray FELs have reported peak electron temperatures approaching 200~eV~\cite{vinko2012creation,Ciricosta:2016}, a value some 20 times higher than the Fermi energy. Performing such calculations using traditional KS DFT is prohibitively expensive. While alternative approaches such as orbital-free molecular dynamics (OFMD) are better suited to treating high temperature conditions, they cannot be used here as we need access to the orbitals directly to study state rebinding.

A recently proposed hybrid scheme by Zhang {\it et al.} seeks to address these challenges by merging aspects of Kohn-Sham DFT with orbital-free DFT~\cite{Zhang2016}. We call this method KS-PWA DFT. Zhang {\it et al.} observe that above a sufficiently high energy, electronic states begin to behave as single plane waves. Therefore the contribution of high energy states to the electron density and the functionals of the Mermin grand potential are simple integrals. These integrals are much faster to calculate than solving the KS equations for these states.
The hybrid scheme then divides the calculations into two regimes, separated by a cut-off energy $\epsilon_C$. Below $\epsilon_C$, where interesting, state-based physics occurs, the Kohn-Sham equations and functionals are still solved in the normal way. Above $\epsilon_C$, the states are assumed to behave as single plane waves, and the integral contributions are added as corrections to the density and to the Mermin grand potential functional calculated from the KS part of the scheme.

This scheme allows us to perform a KS DFT calculation up to arbitrary temperature at the cost of orbital-free DFT while still calculating the full state structure across the lower valence band. We have implemented this scheme in ABINIT v8 for the purpose of the work presented here. As one defines the number of bands to be used in an ABINIT calculation, the cut-off energy is determined by the energy along the uppermost band explicitly calculated.
With this scheme, calculations in primitive unit cells can be performed efficiently at temperatures in excess of 150~eV using only a few hundred bands, whereas it would take many thousands of bands if done using the standard KS DFT method. Equally, plane wave cut-off energies do not need to be as high. The integral corrections required do not add significant computational overhead when compared with standard systems with an equal number of bands, $k$-points, and plane wave energy cut-off.

\section{\label{sec:results}Results}

\subsection{\label{sec:Dim Test Low}Dimensionality in the low-temperature metallic limit}

\begin{figure}
    \centering
    \includegraphics[width=0.48\textwidth,keepaspectratio]{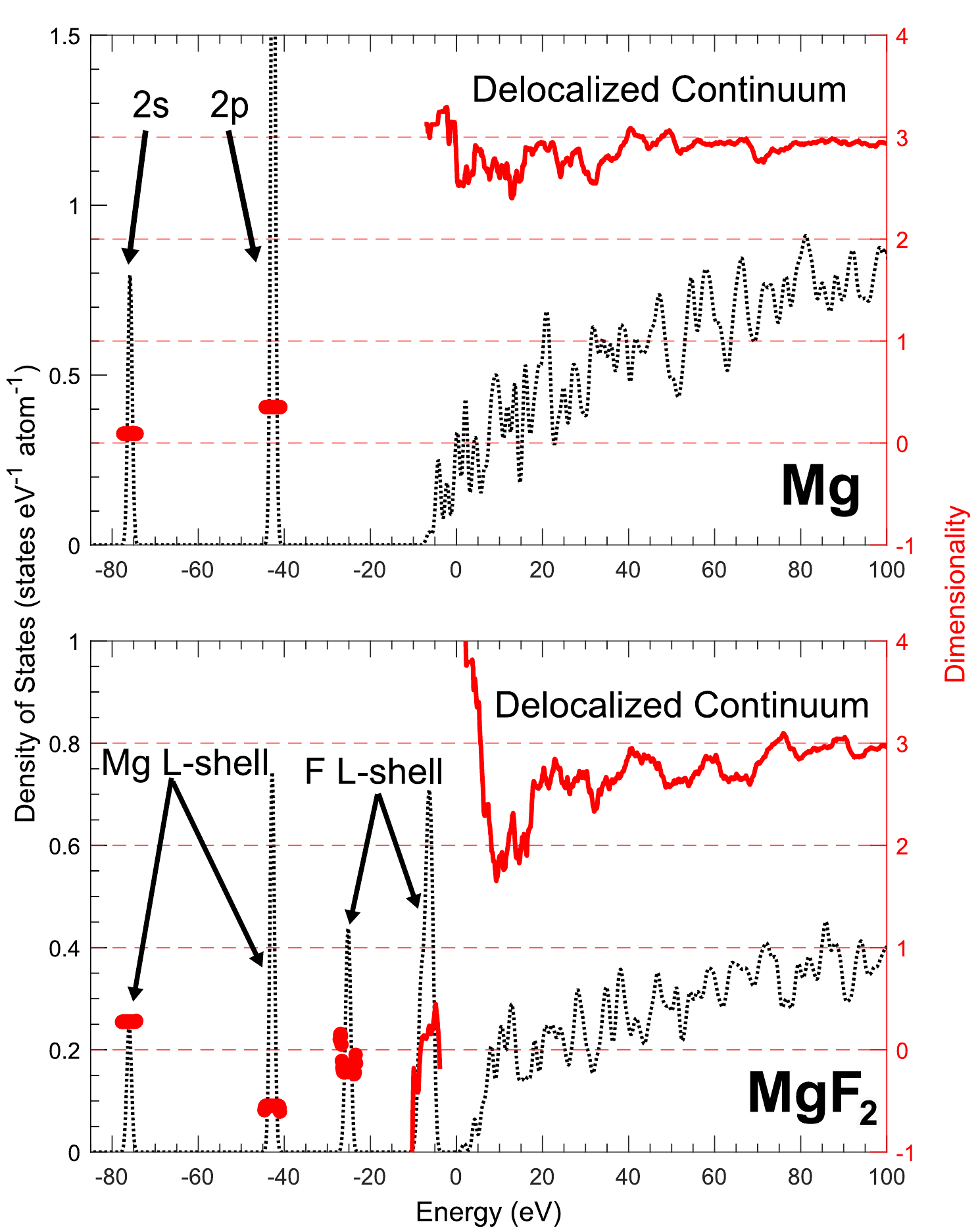}
    \caption{The dimensionality (red solid line) and density of states (black dotted line) of Mg and MgF$_2$ in the ground state. The localized L-shells of the ions have $D < 1$, while the continuum has $D \gtrsim 2$ and $D = 3$ at high energies. This demonstrates the dimensionality correctly predicts localization of states in a simple metal and an ionic compound.}
    \label{fig:LowTempMgMgF2}
\end{figure}

To validate dimensionality as a localization parameter, we apply our dimensionality metric to ground state calculations of a simple metal (Mg) and an ionic compound (MgF$_2$).
The orbitals of isolated atoms are completely localized around the core. However, this behaviour changes at higher densities when an atom can feel the presence of many other neighbouring atoms. In the case of metallic Mg, the $3s$ electrons delocalize to form the conduction band, a continuum that keeps the Mg atoms together. In MgF$_2$, the $3s$ electrons from the Mg ions relocalize around the F sites to fill the F L-shell. In both cases, the K- and L-shell states of the Mg and F ions are localized.
In terms of the dimensionality parameter, the well-localized $2s$ and $2p$ orbitals should have $D < 1$. The simple metal will also have a delocalized continuum, with the continuum having $D=3$ when it has the form of the free electron DOS. The dimensionality, along with the DOS, is plotted in Fig.~(\ref{fig:LowTempMgMgF2}) for the two systems.

In both cases, the L-shells have $D < 1$, indicating these states are localized, as expected. In Mg, the continuum has $D \simeq 3$, showing the dimensionality can identify both localized and delocalized states.
Additional KS states can be included in the calculation without affecting the results when their occupation number is zero. The unoccupied continuum states of MgF$_2$ are included in Fig.~(\ref{fig:LowTempMgMgF2}); they are also delocalized. In both Mg and MgF$_2$, when the eigenenergy of the states is $> 75$~eV the dimensionality of the states is very close to a value of 3, indicating the single plane wave behaviour of the high energy KS states. The dimensionality can concisely identify the localization and delocalization of states for both a simple metal and an ionic compound.

\subsection{\label{sec:CrystalSymmetryEffects}Crystal symmetry effects on localization}

\begin{figure}
    \centering
    \includegraphics[width=0.48\textwidth,keepaspectratio]{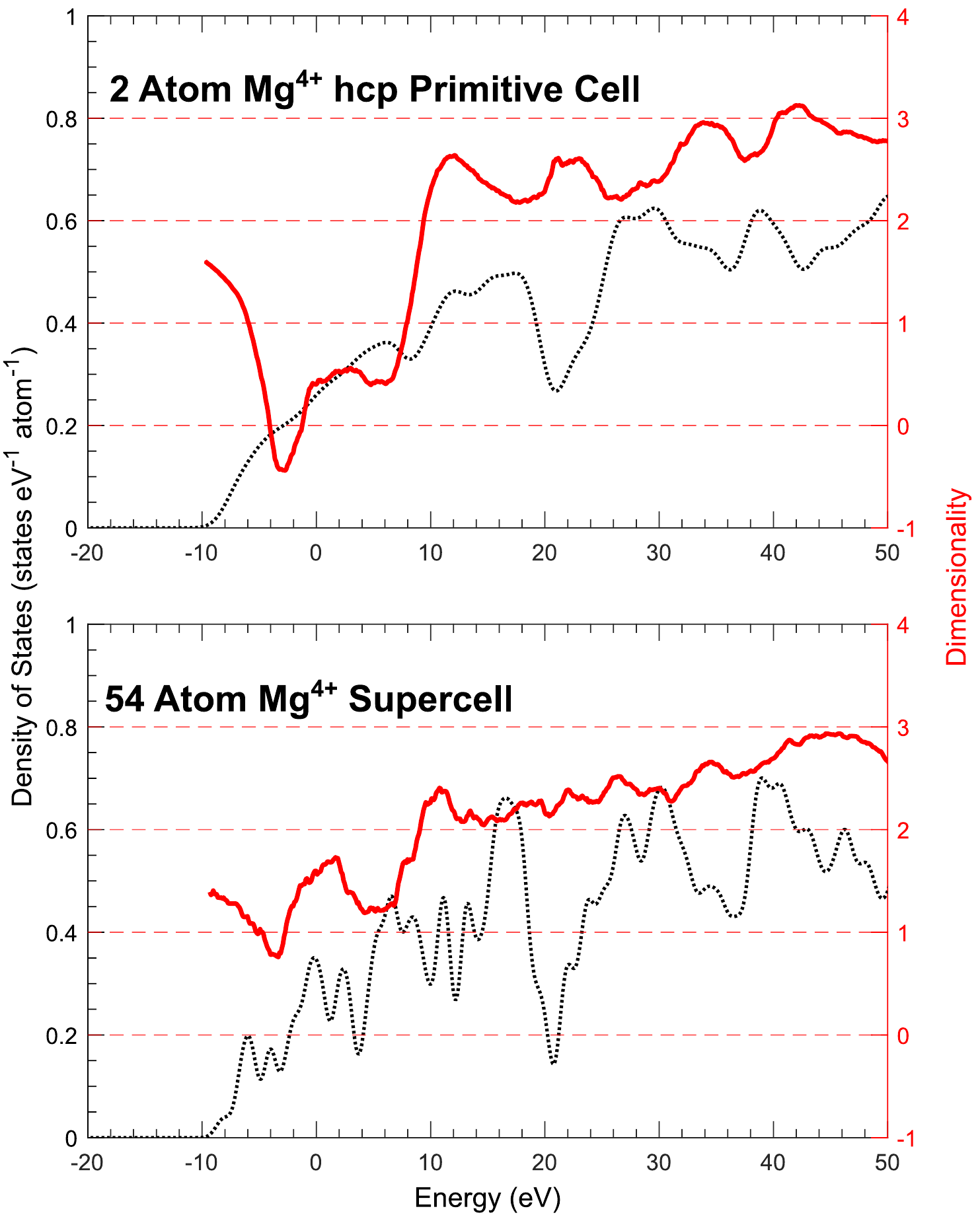}
    \caption{The dimensionality (red solid line) and density of states (black dotted line) for Mg$^{4+}$ in a 2 atom hcp primitive cell and a 54 atom supercell, comparing the effects of crystal symmetry on spatial localization. Crystal symmetry results in an apparent localization effect.}
    \label{fig:PrimVsSuper}
\end{figure}

The derivation of the dimensionality parameter was considered generally for any Bloch state. Therefore, the dimensionality theoretically can be applied to any type of repeating unit cell used in a calculation, including primitive unit cells and supercells. However, there is an interesting localization effect that arises when considering cells with high crystal symmetry, such as the perfect symmetry of primitive cells. In these cases, the perfect crystal symmetry can result in an enhancement in apparent localization, leading to states that appear to be localized where other metrics -- such as the shape of the DOS -- would otherwise suggest.
For example, the top figure in Fig.~(\ref{fig:PrimVsSuper}) plots the dimensionality and DOS for Mg$^{4+}$ ions in the hexagonal close-pack (hcp) primitive cell of Mg. The DOS at the bottom of the continuum shows a square-root dependence on the energy, which is typically associated with free electrons and would therefore indicate delocalization. However, the dimensionality of these states over a broad region has $D \lesssim 1$, suggesting the states are instead localized.
Breaking the perfect crystal symmetry -- such as by considering a supercell with the ions shifted from their perfect lattice positions -- results in noticeably different localization behaviour. The bottom figure in Fig.~(\ref{fig:PrimVsSuper}) the dimensionality is plotted for a 54 atom supercell of Mg$^{4+}$. In this figure, the dimensionality and DOS is averaged over different ion positions from the MD simulations.
As well as being more realistic to experimental measurements, sampling different ion positions allows for only sampling the gamma point in the BZ, which greatly reduces crystal symmetry effects on localization. The atom numbers involved are still relatively small, resulting in a DOS that is much less smooth than the primitive cell case. However, the general features of the DOS between the primitive cell and supercell calculations are the same.
The localization behaviour at the bottom of the continuum has changed dramatically when the crystal symmetry is broken, tending towards delocalized states with dimensionalities $D \gtrsim 1$, as expected. There is a small dip in the dimensionality around $\epsilon = -5$~eV with $D < 1$, which may indicate the supercell used is too small to entirely remove crystal symmetry effects. However, as it is surrounded by delocalized states, it would suggest this region is actually still delocalized.

Supercells still have  crystal symmetry as they are still repeated infinitely via the periodic boundary conditions of the simulation. However, the crystal symmetry is much weaker as it is pushed out to greater distances due to the larger lattice parameters and the random positions of the ions. One would expect then that even larger supercells would further dampen crystal symmetry effects on localization. However, as the calculations performed in this work still require a substantial number of bands, computational memory constrains the number of atoms we can simulate.

\subsection{\label{sec:Heating}Localization of states in hot-dense systems}

As the Al, Mg and MgF$_2$ systems are isochorically heated by the FEL, their L-shell electrons start to thermally ionize, and are stripped from the core of the ion, reducing the overall nuclear screening. As a rule of thumb, the valence states are able to re-localize around individual ions when the strength of the ionized core overcomes the IPD. We now proceed to study the localization of states as the temperature is increased, and the L-shells of F, Mg and Al are increasingly ionized. For convenience we will only discuss integer L-shell ionizations and thus integer charge states (see Fig.~(\ref{fig:ChargeVsT}) for how this translates to temperature for each system), but in reality the ionization is continuous with temperature.

\begin{figure*}
    \centering
    \includegraphics[width=0.95\textwidth,keepaspectratio]{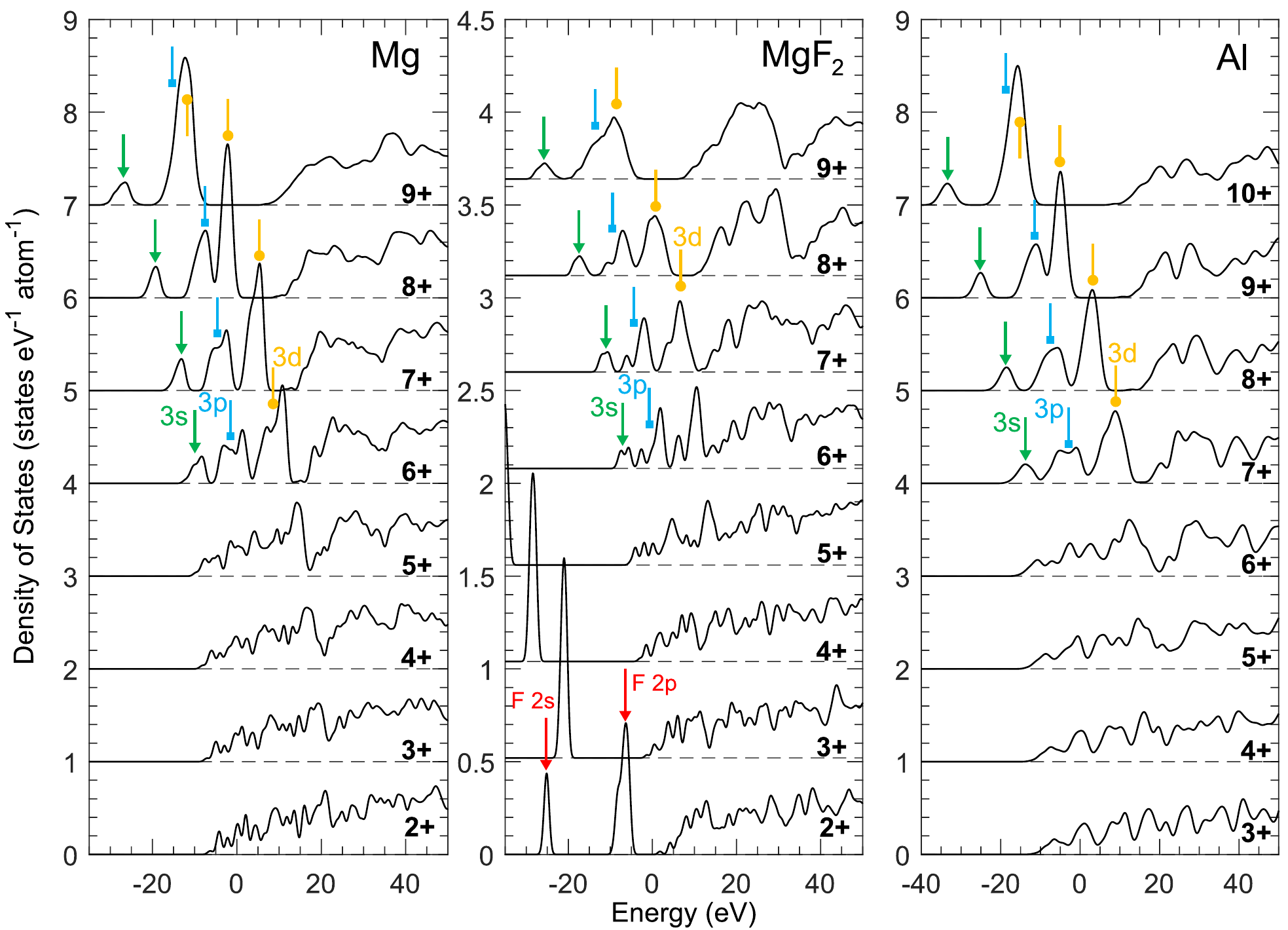}
    \caption{The DOS for Mg, Al and MgF$_2$ as the systems are isochorically heated. The DOS are displaced vertically for clarity, and the energy has been shifted by the Fermi energies of the materials. The charges indicate the charge state of the metal ions in the materials. The three arrows show the location of the $3s$ (green line, arrow), $3p$ (blue line, square), and $3d$ (orange line, circle) orbitals for the metal ions. The presence of these three peaks indicates that the $n=3$ states have started localizing around the metal ions in each material, but does not show the localization mechanism or how well localized these orbitals are.}
    \label{fig:All DOS}
\end{figure*}

Fig.~(\ref{fig:All DOS}) shows the DOS of the supercell calculations of Mg, Al, and MgF$_2$ as the systems are heated. In all cases, three peaks peel off from the continuum DOS; these are the $3s$, $3p$, and $3d$ orbitals of the metal ions in the materials, as determined from the orbital-projected DOS. We note an interesting order of the orbital recombination: the $3d$ orbital has recombined without the $4s$ orbital being present on the ion. It might be expected that the $4s$ orbital would recombine first given that, for an atom, it is typically filled before $3d$ orbital. The $3d$ orbital has also been seen to recombine before the $4s$ orbital in FT-DFT calculations of Na~\cite{Hollebon_2021}. There is therefore a consistent recombination behaviour for the third period metals in which the $3d$ orbital recombines to the metal ions without the $4s$ orbital being present on the ions.
The $3p$ and $3d$ orbitals appear to have merged together at the highest charge states, as a consequence of their eigenvalues getting close together.
In metallic Mg, the three peaks form at a charge state of 6+. Similarly, in Al the peaks are visible for a charge state of 7+. As Mg and Al are both metals and neighbours in the periodic table, it is not surprising they require the same number of L-shell holes for the M-shell to recombine to the metal ions.
In MgF$_2$, the M-shell states also begin to form in the DOS when the Mg ions are at the 6+ charge state. Though, only the $3s$ and $3p$ orbitals are clearly forming -- the peaks that will form the $3d$ orbital remain largely merged with the rest of the continuum. The M-shell states are fully formed when the Mg ions have a 7+ charge state.
In all cases, the occupation number of the M-shell orbitals is small, with around or less than one electron shared between all three orbitals on each ion.
Peaks forming in the DOS intuitively suggests that states are localized to some extent, and the dimensionality parameter allows us to examine this in more detail. In particular, it can be used to identify how the M-shell states relocalize to the Mg ions.

\begin{figure*}
    \centering
    \includegraphics[width=0.95\textwidth,keepaspectratio]{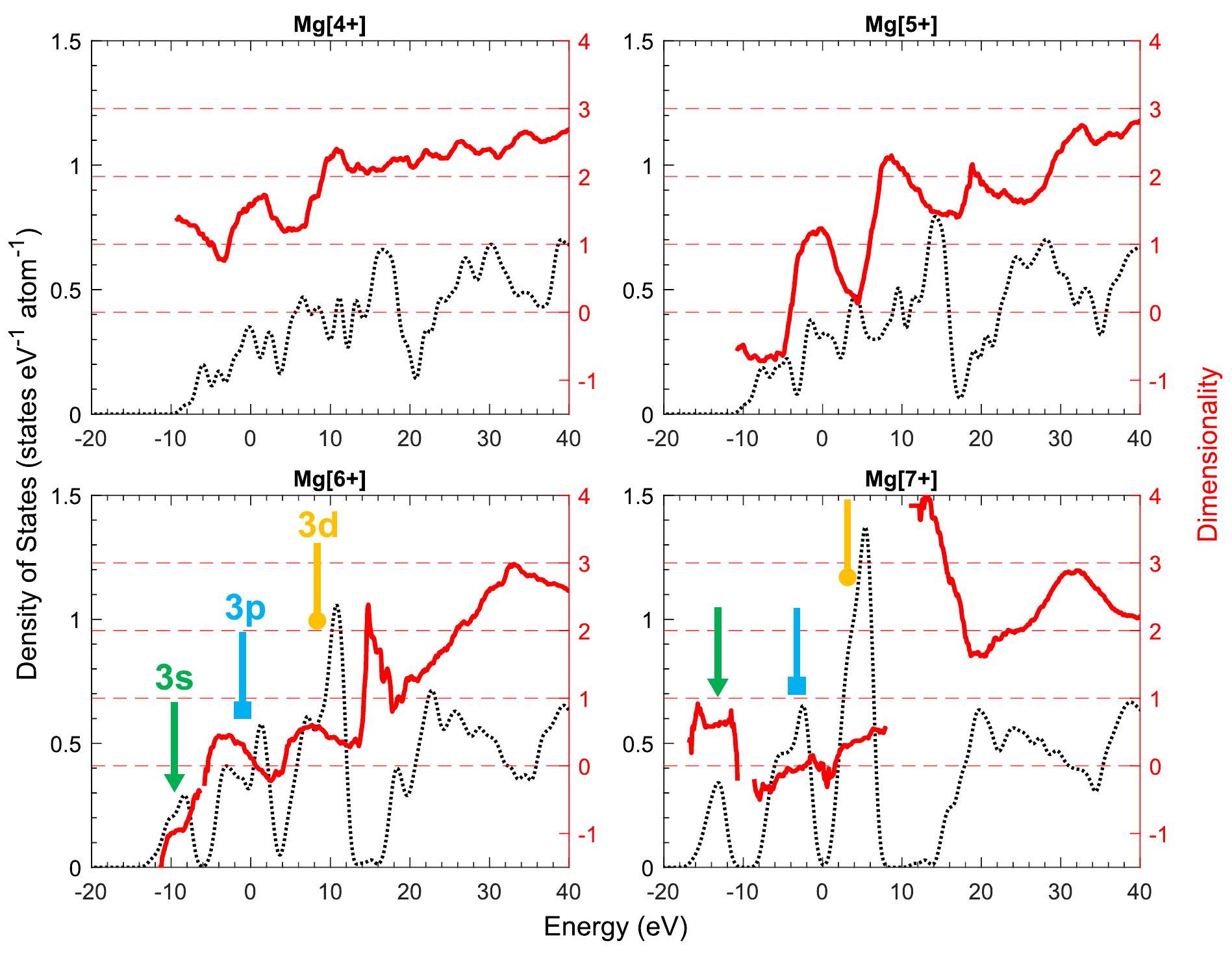}
    \caption{The dimensionality (red solid line) and density of states (black dotted line) for hot Mg with a mean charge state between 4+ and 7+. The three peaks that develop correspond to the M-shell. The dimensionality shows the state relocalization occurs while the M-shell recombines to the Mg sites.}
    \label{fig:Mg 4-7}
\end{figure*}

The dimensionality and DOS for Mg with charge states 4+ to 7+ is illustrated in Fig.~(\ref{fig:Mg 4-7}). For Mg$^{4+}$, the states near the continuum edge are in the process of relocalizing around the Mg ions, with the dimensionality of the bottom of the continuum being $D < 3$. At higher energies, the continuum still has $D = 3$. In Mg$^{5+}$, the bottom of the continuum at energies $\epsilon < -2.9$~eV has $D < 0$, indicating there are states that have relocalized. An equivalent primitive cell calculation had this region correspond to a peak which was a localized quasi-$3s$ orbital, with the majority angular momentum contribution from $l=0$, and additional contributions from other angular momentum channels. Between $\epsilon = $ -2.9 and 5.7 eV, $D \lesssim 1$, indicating there are localized states -- the primitive cell calculation reveals this to be a quasi-$3p$ orbital. The quasi-orbitals are not distinguishable in the supercell calculations as the quasi-orbitals are broad, and their eigenvalues strongly depend on the ion positions. Therefore, due to sampling over many shifted ions, the quasi-orbitals blur together and are indistinguishable from the continuum in the DOS of the supercell calculation.
In Mg$^{6+}$, the $n=3$ orbitals have fully relocalized, and the three peaks have $D < 1$. In the IPD picture, the highly-ionized core has overcome the IPD and the M-shell has recombined to the Mg ions. At higher energies, there is still a delocalized continuum with $D\approx3$.

\begin{figure*}
    \centering
    \includegraphics[width=0.95\textwidth,keepaspectratio]{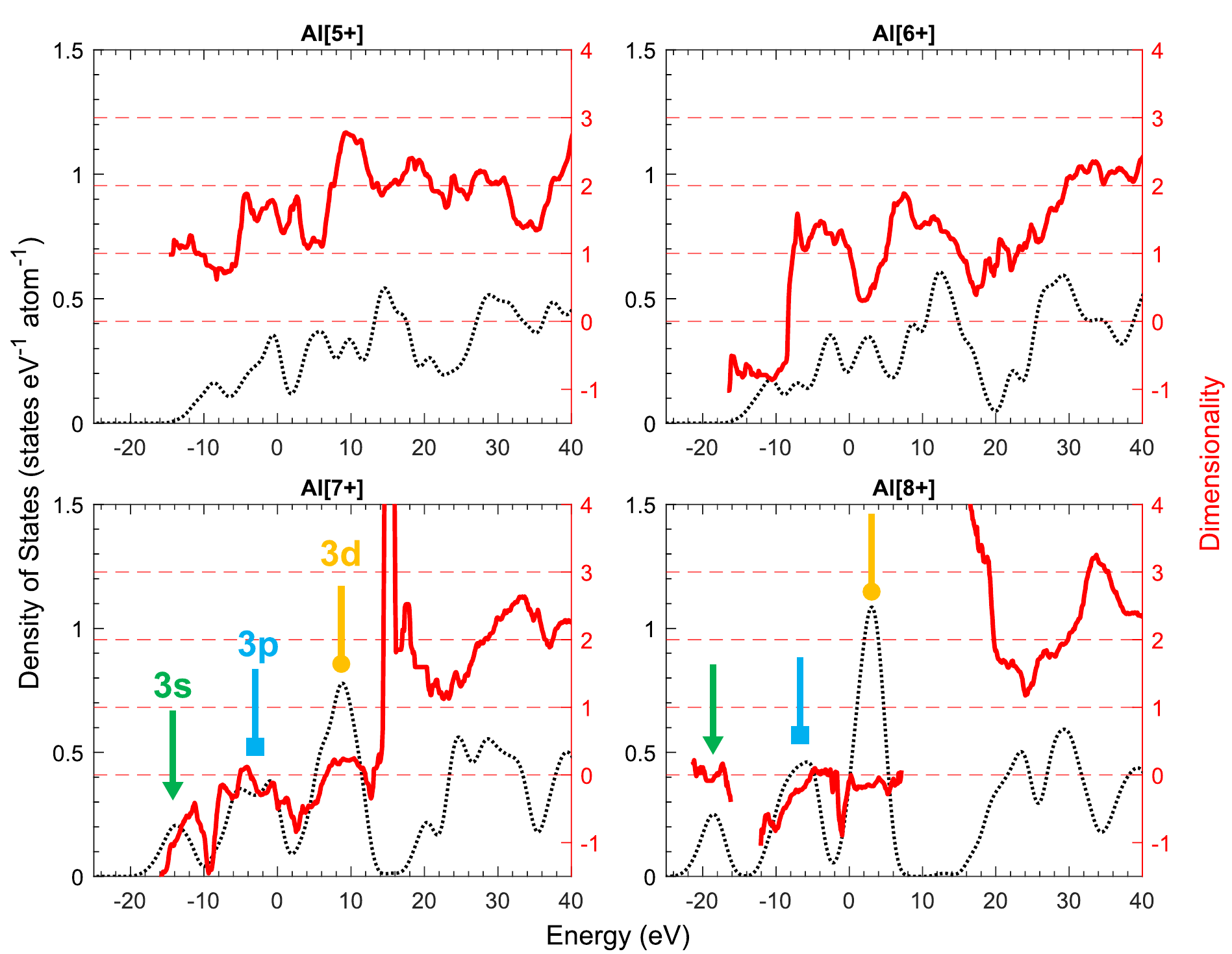}
    \caption{The dimensionality (red solid line) and density of states (black dotted line) for hot Al with a mean charge state between 5+ and 8+. The three peaks that develop correspond to the M-shell. The dimensionality shows the state relocalization occurs while the M-shell recombines to the Al sites.}
    \label{fig:Al 5-8}
\end{figure*}

The relocalization of the Al M-shell is plotted in Fig.~(\ref{fig:Al 5-8}). The relocalization behaviour is very similar to Mg, requiring the same number of L-shell electrons to be ionized in order for the M-shell states to relocalize. We note a sharp jump in the dimensionality at $\epsilon \approx 15$~eV in Al$^{7+}$, indicating a strong sensitivity of the continuum states on the lattice parameter immediately as the M-shell relocalizes.

\begin{figure*}
    \centering
    \includegraphics[width=0.9\textwidth,keepaspectratio]{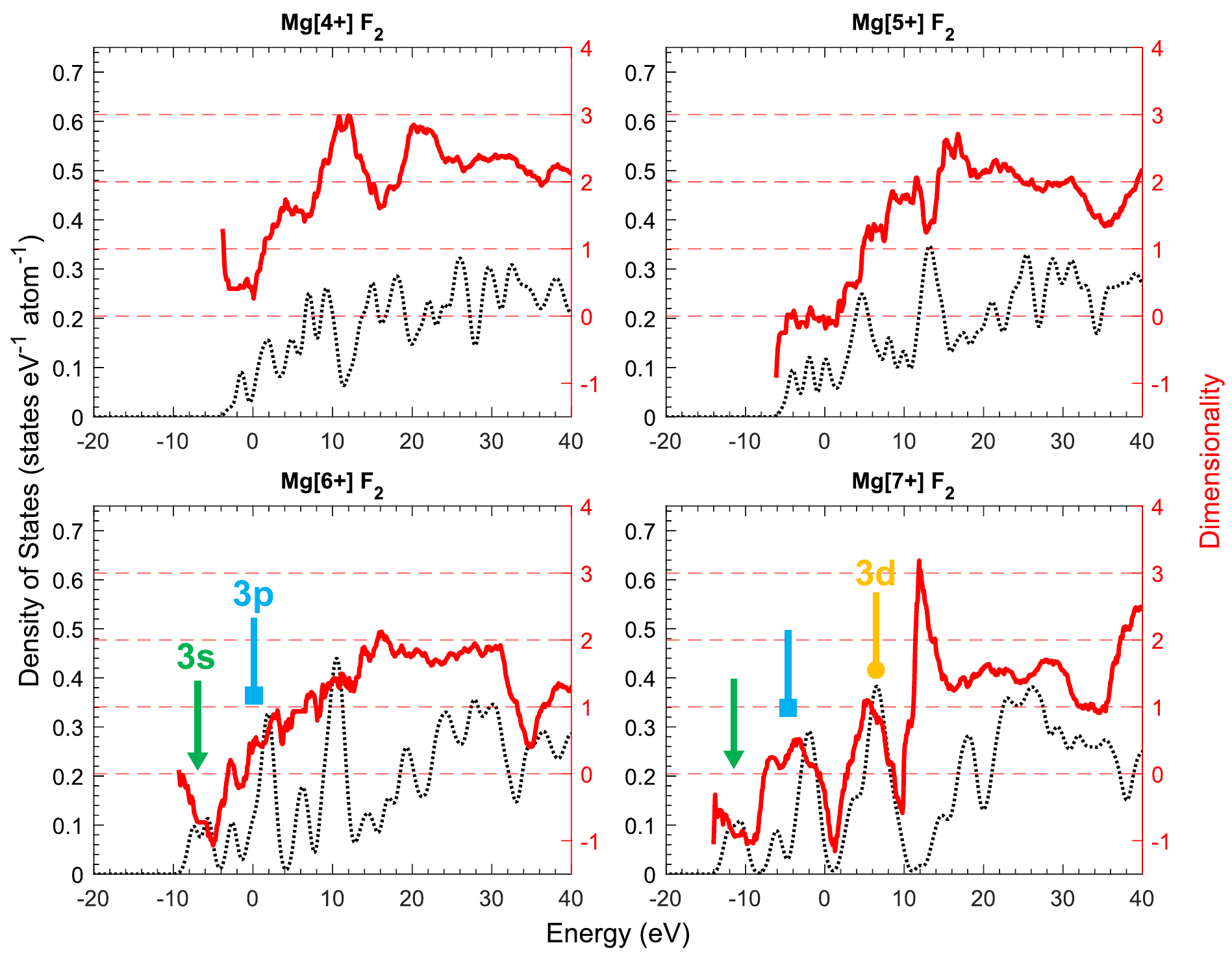}
    \caption{The dimensionality (red solid line) and density of states (black dotted line) for hot MgF$_2$ for Mg charge states of 4+ to 7+. The three peaks that develop correspond to the M-shell of the Mg ions. An additional broad, structurally-sensitive feature is also present just above the continuum edge for charge states $\rm{Mg^{7+}F_2}$ and above.}
    \label{fig:4-7 MgF2}
\end{figure*}

We also examine the relocalization process in MgF$_2$. The density of Mg ions in MgF$_2$ is lower ($3.1 \cross 10^{22}$ cm$^{-3}$) than in Mg ($4.3 \cross 10^{22}$ cm$^{-3}$). For IPD models, a lower ion density results in a lower IPD energy. However, the F ions are easier to ionize than the Mg ions, so as Mg and MgF$_2$ are thermally-ionized, the free electron density in MgF$_2$ will be higher than in Mg at the same temperature. The F ions act as donors to increase the electron density.
Despite the competing IPD effects from the electron and ion density, IPD models predict that the IPD energy is greater in MgF$_2$ than in Mg. In general, it was found that IPD models are poor at predicting CL in compounds \cite{Ciricosta2016-nx}. MgF$_2$ is therefore used to compare the differences in relocalization between a simple metal and an ionic compound, and the effects of additional electron density from donor ions on the relocalization process of thermal ionization.
The relocalization of the Mg M-shell is plotted in Fig.~(\ref{fig:4-7 MgF2}). It is still a smooth process, and broad M-shell states relocalize around the Mg ions, however there are notable differences compared to Mg.
First, the charge state of the Mg ions required to fully relocalize the M-shell is higher in MgF$_2$ than in Mg; 7+ versus 6+, respectively.
Additionally, the M-shell states do not relocalize together at the same charge state. Instead, the $3s$ and $3p$ orbital have relocalized in $\rm{Mg^{6+}F_{2}}$, but the $3d$ relocalizes at $\rm{Mg^{7+}F_{2}}$. The difference in the relocalization behaviour of the M-shell compared to Mg is attributed to differences in the electron density around the Mg ions in the two materials.
Immediately after the continuum level, there is a feature from $\epsilon = $ 15~eV to 38~eV where the dimensionality is approximately constant with $1 \lesssim D < 2$. This feature is sensitive to the positions of the ions, and consists of contributions from both the Mg and F ions and many different angular momenta. Nonetheless, on average, it appears to be delocalized like the rest of the continuum. At higher energies, $D \approx 3$ as expected.

There are some similar behaviours seen in a all three materials. Firstly, the dimensionality indicates that states relocalize at the bottom of the continuum before there are any clear orbital features. Calculations using primitive cells of the materials indicates that these correspond to quasi-M-shell orbitals. However, as the widths and eigenvalues of the orbitals are strongly dependent on the structure of the system, due to the different electron densities surrounding the ions in different positions, these orbitals merge and blur together. This would indicate that despite the presence of localized states, they would be indistinguishable from the continuum.
When the metal M-shell states do become distinct, they are significantly broader than the isoenergetic metal ion L-shell states. For example, the full width of the $3p$ orbital for $\rm{Mg^{6+}}$ and $\rm{Mg^{7+}}$ is $\sim 10$ eV. Despite the relative broadness of the M-shell states, they are fully relocalized as they recombine to the ions. 
The large width of $3p$ orbital is in agreement with the observation of the very broad K$_{\beta}$ transitions seen in Fig.~(\ref{fig:Mg MgF2 Spec}). While the K$_{\beta}$ transitions are broader than the $3p$ orbitals, FT-DFT only considers a mean static system, so it neglects other contributions such as configurational broadening. Numerous broad, overlapping features contribute to the overall large width of the K$_{\beta}$ transitions.

The continuum states have $D>1$, showing that we still have a delocalized continuum.
There is a sharp increase in the dimensionality at the bottom of the continuum once the M-shell states have relocalized. We note a similar feature is seen in MgF$_2$ in Fig.~(\ref{fig:LowTempMgMgF2}), where states immediately above the Fermi level (the zero energy in the plot) also show large dimensionalities ($D \approx 4$) that reduce to $2 \lesssim D \lesssim 3$ as the energy of the states increases. These states at the bottom of the continuum are delocalized, but are not a free-electron gas. The high dimensionality would suggest they are sensitive to the structural conditions of the system, with separating the ions resulting in an increased delocalization.
At higher energies the dimensionality settles to $D=3$ as expected.

\section{\label{sec:discussion}Ionization Potential Depression}

\begin{figure}
    \centering
    \includegraphics[width=0.48\textwidth,keepaspectratio]{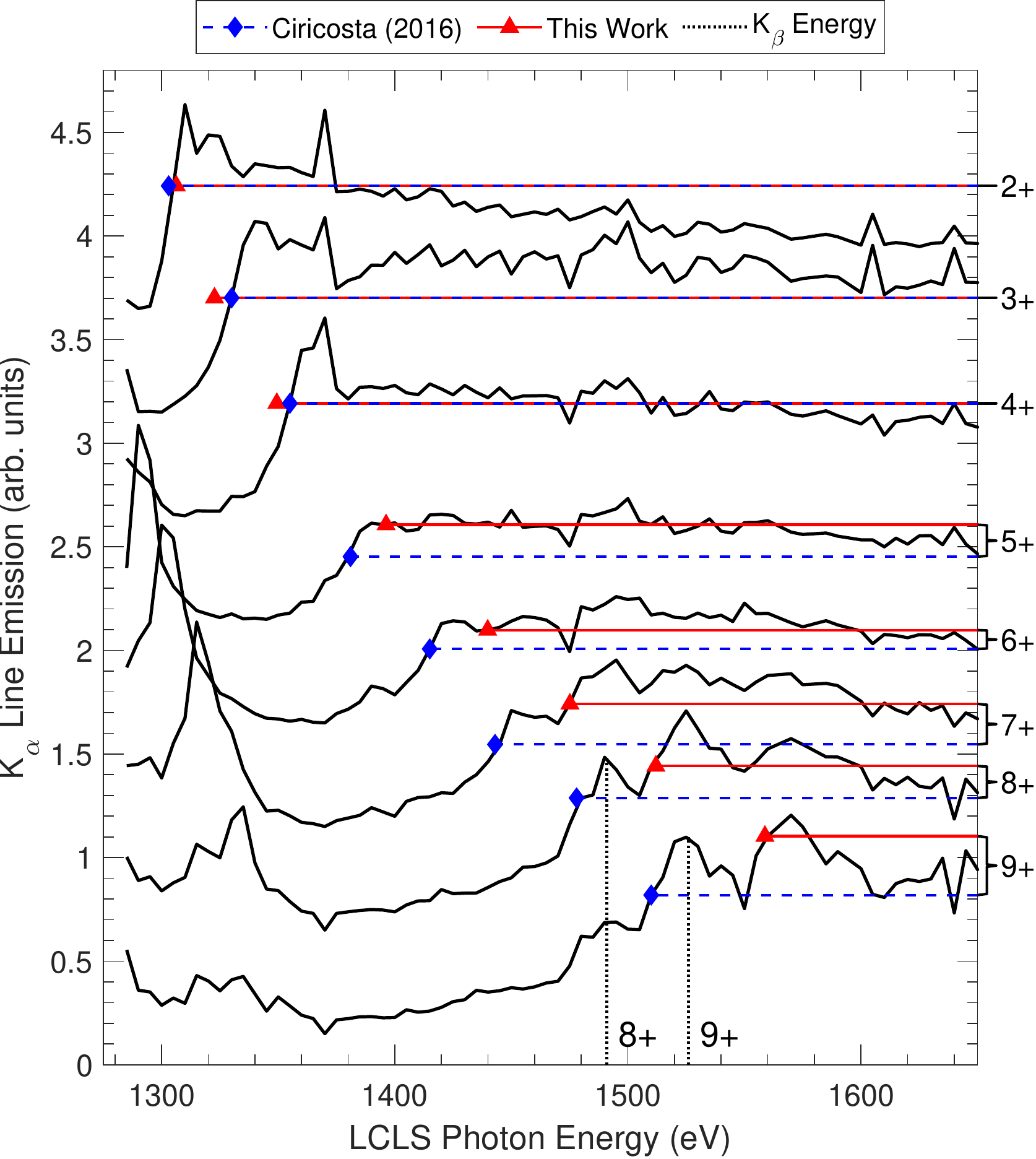}
    \caption{Lineouts for the emission intensity of the K$_{\alpha}$ lines of Mg at different charge states, reproduced from Ciricosta (2016) \cite{Ciricosta2016-nx}. The horizontal dashed blue lines indicates the K-edge energies inferred in Ciricosta (2016). There is a $\pm 5$~eV uncertainty due to the resolution of the measured FEL photon energy. The vertical black dotted lines indicate the measured Mg K$_{\beta}$ lines from the peaks in Fig.~(\ref{fig:Mg MgF2 Spec}). The horizontal solid red lines indicate the K-edge energies extracted using FT-DFT. For the 2+ to 4+ charge states, the Ciricosta (2016) edges overlap with the FT-DFT edges.}
    \label{fig:OC16}
\end{figure}

In the experiment presented in ref.~\cite{Ciricosta2016-nx}, the x-ray FEL was used to irradiate Mg foils over a range of x-ray photon energies, selectively ionizing the K-shell in increasingly high ion charge states. Only when the driving photon energy exceeds an ionization threshold, or matches an excitation energy, can a $1s$ core hole be created. These core-hole ions radiatively recombine predominantly via the K$_{\alpha}$ emission channel, providing a spectral signature of the charge states excited during the irradiation process. As there are eight electrons in an L-shell, eight K$_{\alpha}$ lines can be observed. The experimental emission intensity of these eight lines is plotted against the driving FEL photon energy in Fig.~(\ref{fig:OC16}), reproduced from ref.~\cite{Ciricosta2016-nx}. Clear edges are observed for the K$_{\alpha}$ lines, indicating a threshold energy required to excite a $1s$ electron from the K-shell. In the absence of a bound M-shell, these thresholds represent the K-edges of the dense Mg system, and these are marked on the figure with blue lines. The difference in the measured ionization threshold in the solid-density system and the ionization energy of isolated Mg atoms gives an experimentally measured IPD energy.

We proceed to extract the IPD energy from our DFT simulations, taking into account the dimensionality metric described in the previous section, and compare these results with the ref.~\cite{Ciricosta2016-nx} experimental data shown in Fig.~(\ref{fig:OC16}).
Vinko~\textit{et al.}~\cite{Vinko2014-fs} detailed two methods for extracting K-edge energies from DFT calculations. One is to take directly the energy difference between two DFT calculations, one with and one without a K-shell hole, while accounting for the additional energy required to thermalize the ionized electron. The second is to take the difference between the energy of the $1s$ orbital and the continuum edge. Both methods produce similar results. For the work here we chose the second method, as it is more convenient when treating hot systems where the thermal energy in the valence electrons is considerable. It also makes it easier directly to include the dimensionality measure in the edge evaluation.

The continuum edge energy $E_c$ is calculated in the extended system using a plane wave DFT calculation, as this allows us appropriately to model the extended distribution of free electrons. In IPD models, the continuum edge is the boundary between an electron being bound or free. Following this definition, $E_c$ is placed at the separation between localized and delocalized states. As previously noted, this separation can lie in a continuous region of the supercell DOS. The DFT calculations therefore suggest some localized states cannot be easily be distinguished from the delocalized continuum. The energy of the $1s$ orbital cannot be extracted directly from the same calculation as a pseudopotential is needed to represent the tightly bound $1s$ states in the plane wave DFT approach. Instead, the $1s$ energy can be found relative to the $2p$ state by computing the K$_{\alpha}$ energy $E_{\rm K_{\alpha}}$ for the charge state of interest using the atomic DFT code~\textit{Atompaw} from which the frozen-core pseudopotential was constructed. This energy can then be subtracted from the $2p$ orbital energy $E_{2p}$ found in the plane wave DFT calculation to yield the binding energy of the $1s$ state. The K-edge energy $E_{\rm K}$ is therefore computed as $E_{\rm K} = E_{c} - \left(E_{2p} - E_{\rm K_{\alpha}} \right)$. We plot the results of this calculation for the eight Mg K$_{\alpha}$ lines in Fig.~(\ref{fig:OC16}) with the red triangles. We see that for the first three charge states our K-edge predictions match the thresholds observed in the data, but start to differ for the highest five states. This is because the DFT simulations predict a relocalization of the M-shell for Mg ions with more than four holes in the L-shell. This M-shell relocalization process is depicted in Fig.~(\ref{fig:Mg 4-7}) across the relocalization region for charge states between 4+ and 7+.
We note that if one takes $E_c$ for Mg$^{5+}$ as the bottom of the continuous region of the DOS, the DFT simulations agree well with the observed data. This further highlights the difficulty in distinguishing the delocalized continuum from close lying localized states, as well as demonstrating that the continuum generally has a more complex localization behaviour beyond the free-electron gas.
Strictly speaking, once the M-shell has relocalized we can no longer identify the charge state of the ion using only the number of electrons in the K- and L- shells as there may be bound M-shell spectator electrons. However, we choose to retain this Ne-core based nomenclature to keep a consistent notation, and also note that the high temperatures and low binding energies of the M-shell mean that this will not substantially alter the energetics of the system.

The observation of M-shell states in highly charged ions in the DFT simulations allows us to identify the $1s \rightarrow 3p$ excitation transitions in the experimental data of Fig.~(\ref{fig:OC16}), marked in black. These structures in x-ray absorption (and subsequent K$_{\alpha}$ emission) are produced by an excitation into an M-shell state, rather than an ionization event into the continuum. While these excitation peaks are prominent for states 8+ and 9+, the process is far less obvious for states 6+ and 7+, neither of which display particularly notable features. The 5+ state presents no notable feature. In contrast, some prominent peaks do not require us to reevaluate the position of the K-edge. For example, the peak at 1525~eV observed in state 8+ can be understood as a combination of a) the ionization of the 8+ ion; and b) a resonant K$_{\beta}$ excitation in the 9+ ion, followed by a collisional recombination event in the L-shell before the K-shell is filled via radiative K$_{\alpha}$ emission in a now lower charge state 8+. Such ultrafast L-shell collisional dynamics within the duration of the $1s$ core hole lifetime has been studied previously for both ionization~\cite{Vinko2015} and resonant K$_{\alpha}$ excitation~\cite{van2018clocking}, but not for resonant K$_{\beta}$ excitation as seen here. Nevertheless, it is unsurprising that a similar process should occur for resonant K$_{\beta}$ excitation once the M-shell relocalizes.

\begin{figure}
    \centering
    \includegraphics[width=0.48\textwidth,keepaspectratio]{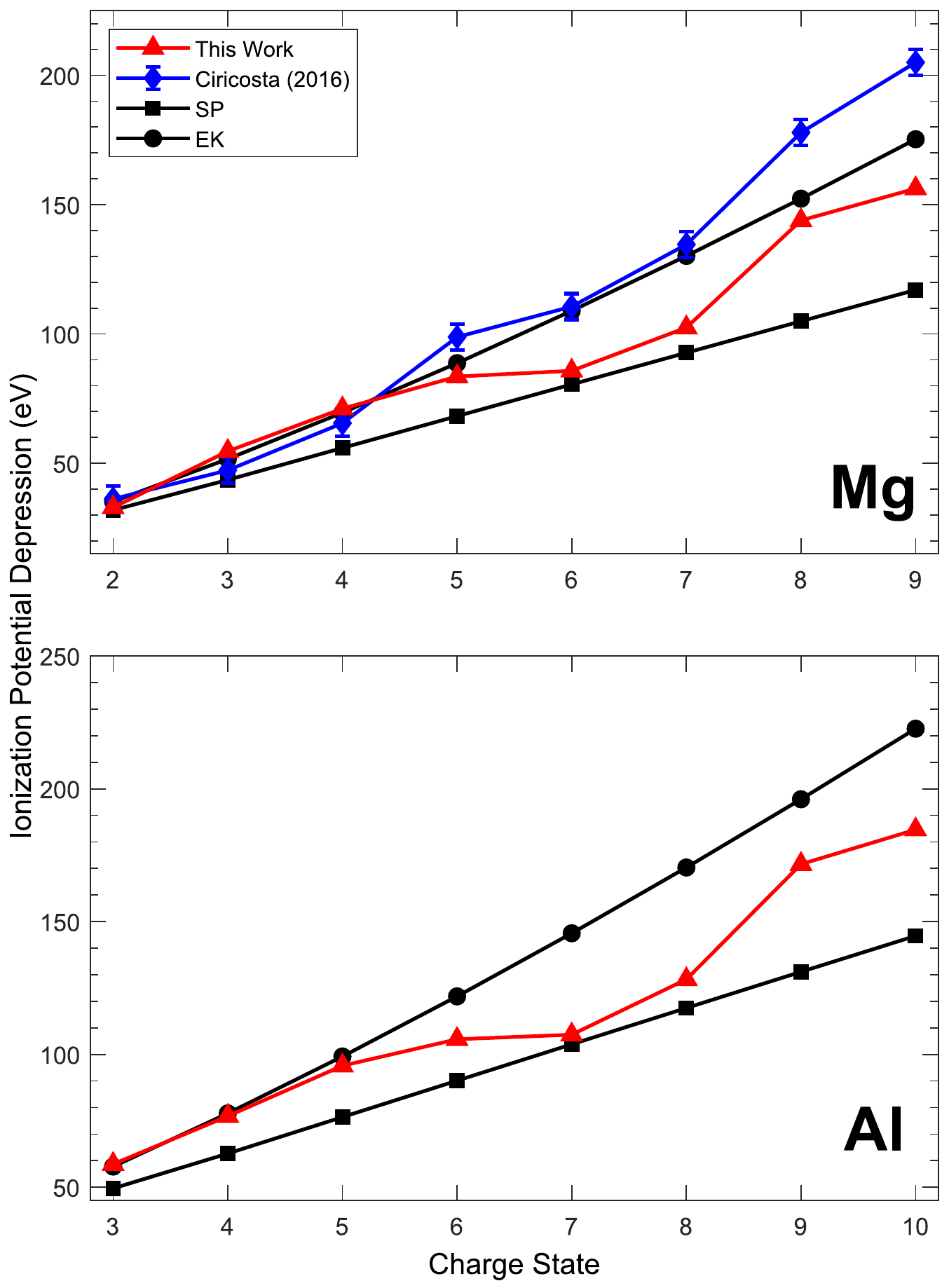}
    \caption{IPD levels for different charge states of Mg and Al, inferred from various schemes: the FT-DFT calculations in this work (red triangles), the EK model (black circles), the SP model (black squares), and experimental data published in Ciricosta (2016)~\cite{Ciricosta2016-nx} (blue diamonds).}
    \label{fig:MetalIPD}
\end{figure}

The IPD can be extracted from the DFT calculations by taking the difference of the $E_{\rm K}$ energy and the energy required to ionize the $1s$ orbital in an isolated ion with the same charge state. We plot these energies in Fig.~(\ref{fig:MetalIPD}), alongside the experimental measurements of ref.~\cite{Ciricosta2016-nx}, and the predictions from the analytical SP and EK models. As expected, there is agreement between the results of this work and the IPD values found by ref.~\cite{Ciricosta2016-nx} for the first three charge states, but beyond that the predictions diverge due to the emergence of relocalizing states and a rebinding M-shell. We note that the IPD values agree with the EK model only for the first few charge states, with the EK model overestimating the IPD at high ionization. There is no systematic agreement with the predictions of the SP model which generally underestimates the IPD. Overall, an IPD energy between the EK and SP models at high charge states is qualitatively consistent with the results of other previous experiments \cite{Ciricosta2013PRL,Hoarty2013,Ciricosta2016-nx,hansen2017changes} in similar plasma conditions. There is also broad agreement between our results and other theoretical studies~\cite{son2014quantum,crowley2014continuum,calisti2015ionization,zeng2022ionization}, but it is worth mentioning that the underlying assumptions are quite different between models, and few explicitly treat the rebinding of M-shell states in the continuum.

Given the proximity of Al and Mg on the periodic table it is unsurprising to note a similar trend in the predicted level of IPD in Al. As in Mg, Al also shows the M-shell relocalizing at higher charge states, leading to a more complicated IPD trend as a function of charge state. Again, the IPD values follow the EK model well until the M-shell rebinds, after which the IPDs lie between the predictions of the SP and EK models. Previous experimental work investigating the IPD in Al was only able to extract the IPD for the first 5 charge states~\cite{Ciricosta2013PRL}. Our calculations agree with these results for all but the highest charge states. The difficulty in extracting experimental IPD values for the highest charge states in Al is consistent with the predicted complex IPD behaviour from our calculations, which cannot be reproduced by simple IPD models. This severely hinders our ability to model experiments using time-dependent atomic kinetics simulations, on which much of the experimental IPD work is based.

We can also compare our IPD predictions with previous DFT-based results of Vinko {\it et al.}~\cite{Vinko2014-fs}, which generally showed good agreement with the predictions of the EK model. The method described here is similar to ref.~\cite{Vinko2014-fs} in many ways, but differs in how the continuum is identified. Ref.~\cite{Vinko2014-fs} assumed that the K-edge energy (and thus the IPD) could be found by comparing two DFT simulations for various charge states with and without a K-shell core hole. The system was always assumed to be charge-neutral, and any electron removed from the K- and L-shells was placed in the continuum and equilibrated. The energy difference calculated in this way therefore always assumed that the valence states are part of the continuum, and the K-edge is effectively defined as the lowest possible energy that allows for the excitation of a $1s$ electron. This method can successfully predict all the excitation thresholds, but is only a reliable way to find the K-edge if no states fully rebind in the continuum. The assumption of free valence electrons was based on the observation of relatively broad M-shell features in the DOS, which are normally indicative of unbound states. However, as we have seen in our IPR and dimensionality analysis here, the behaviour of the valence band is complex and the shape of the DOS alone is not always a reliable indicator of whether a state is bound or free. While the distinction between an excitation threshold and a K-edge is not critical for some plasma properties (eg. the energetics), it does matter in the context of x-ray spectroscopy and IPD. For such investigations the more detailed analysis presented here is required.

Another important plasma effect that is included self-consistently in the DFT calculations, but is typically excluded in atomic-kinetics and IPD calculations, are level shifts due to the surrounding plasma particles, through so-called plasma polarization shift (PPS) \cite{GriemPlasmaSpectroscopy}. PPS arises due to the surrounding plasma particles perturbing the wavefunctions of an ion's orbitals, with the perturbation increasing with the principle quantum number of an orbital \cite{renner1998experimental}. Experimental evidence of PPS for a number of elements and plasma densities \cite{renner1998experimental,saemann1999isochoric,woolsey2000experimental,eidmann2003k,renner2006spectral,khattak2012evidence,sobczuk2022plasma} indicates it red-shifts transition lines. In other words, the binding energy of outer orbitals increases more relative to inner orbitals. Atomic-kinetics codes generally rely on atomic datasets in quantifying the binding energies of different levels. An additional explanation for why the IPD models differ from the DFT calculation is failing to account for changes in the binding energy of orbitals by the surrounding plasma. As this is occurring due to changes in the orbital wavefunction, it is not inconceivable that an orbital may remain bound if a perturbation were to keep it below the IPD-adjusted continuum level. PPS should therefore be accounted for in future collisional-radiative atomic-kinetics codes or in models of IPD.

\section{\label{sec:conclusion}Summary}

We have presented the first experimental observation of K$_{\beta}$ emission in highly ionized solid density Mg plasmas driven by an intense x-ray FEL. Our results indicate that the M-shell rebinds in highly charged Mg ions. This result is consistent with some popular IPD models, but contradicts several others. While the IPD and the need to distinguish between bound and free states in a dense plasma is an artificial construct from a physics standpoint, it is essential to many plasma physics simulations, including collisional-radiative modelling. As such, finding a robust method to determine the IPD in dense plasmas without the need for empirical parameters remains of considerable practical interest.

To interpret the experimental results and place the modelling of the IPD on a firm footing we introduced a theoretical model for the calculation of the IPD from first principles, based on finite-temperature density functional theory. Within this framework we describe a localization parameter, the dimensionality, that allows us reliably to evaluate whether a valence state should be considered bound or free, based on its spatial localization. We apply the model to x-ray isochorically heated Al, Mg and MgF$_2$, and find that the $n=3$ states relocalize at high ionization in all three systems. The relocalization process is seen to be weaker in MgF$_2$ than in Mg owing to the change in environment around the Mg ions due to the presence of additional delocalized electrons from the F ions. This is consistent with experimental observations. 
While our approach builds on the previous DFT work of Vinko {\it et al.}~\cite{Vinko2014-fs}, we show the importance of the dimensionality metric in distinguishing between excitation and ionization thresholds. While essential to compute a well-defined IPD energy, the concept of boundness of a state is not trivial to define for a dense plasma within the general framework of Kohn-Sham DFT.

In comparing our predictions with simple analytical IPD models, we find agreement with the EK model only for the first four lower charges states in all three systems. For higher charge states we observe considerable complexity in the relocalization of M-shell states, a process which cannot in general be reproduced by simple analytical models. In particular, for these higher ionisation states the EK model overestimates the IPD, while the SP model systematically underestimates it. Based on this observation we have revisited the interpretation of recent experimental results from Ciricosta {\it et al.}~\cite{Ciricosta2016-nx}, and have shown how observed spectral features in the measured intensity of K$_{\alpha}$ emission as a function of x-ray photon energy can be attributed to the resonant pumping of the K$_{\beta}$ transition for the higher ionisation states.

Thus, as an overarching conclusion, we find that the original experimental work performed at LCLS that identified the EK model as providing a better description of the IPD than the SP model in Al~\cite{Ciricosta2013PRL} is justified, given that particular work provided IPD values for charge states only up to 7+.  However, the current observation of rebinding of the M-shell for higher ionisation states is also consistent with the work of Hoarty {\it et al.}~\cite{Hoarty2013}. These findings further illustrate the inherent difficulties in attempting to use simplified models of the IPD in atomic kinetics calculations: as the authors of ref.~\cite{Ciricosta2013PRL} stated in their paper {\it ``The reasons underlying the success of the EK over the SP model are far from clear: both the SP and the EK models are simple, semi-classical models, ultimately both unlikely to capture fully the complex physics of atomic systems embedded within dense plasma environments over wide ranges of plasma conditions and charge states.''}

While the DFT-based model presented here provides new insight into the mechanisms of continuum lowering, at present the model remains too computationally cumbersome to be implemented in-line in atomic kinetics suites or in plasma opacity codes over multiple charge and atomic states. For this use, simpler analytical models are still required. Nevertheless, we trust that our approach, alongside the comparison with available experimental data, can provide a valuable framework to evaluate models for use across a wide range of plasma temperatures, densities and ionizations.

\section*{Acknowledgements}
T.G., J.S.W. and S.M.V. acknowledge support from AWE via the Oxford Centre for High Energy Density Science (OxCHEDS).
A.F. acknowledges support from the UK STFC XFEL Hub.
T.C. and S.M.V. acknowledge support from the Royal Society.
M.F.K., S.R., R.R., J.S.W. and S.M.V. acknowledge support from the UK EPSRC under grants EP/P015794/1 and EP/W010097/1.
G.P.-C. acknowledges support from Spanish Ministry of Science and Innovation under Research Grant No. PID2019-108764RB-I00.
S.M.V. is a Royal Society University Research Fellow.

\bibliographystyle{apsrev4-2}
\bibliography{refs.bib}

\end{document}